\documentclass[letterpaper,11pt]{article}
\pdfoutput=1
\usepackage{amssymb}
\usepackage{amsmath}
\usepackage{amstext}
\usepackage{graphicx,epsfig}
\usepackage{epsfig}
\usepackage{verbatim} 
\usepackage{fancybox}
\usepackage{color}
\usepackage{ulem}
\usepackage{enumitem}
\usepackage{subfigure}
\usepackage{bbm}
\usepackage{parskip}
\usepackage{dsfont}
\usepackage[numbers,sort&compress]{natbib}
\usepackage[body={6.5in, 9in}, right=1in, top=1in]{geometry}

\newcommand{\Comment}[1]{{}}
\definecolor{darkblue}{rgb}{0.15,0.35,0.55}
\definecolor{reddish}{rgb}{0.65, 0.2, 0.2}
\usepackage[linktocpage=true]{hyperref}
\hypersetup{
colorlinks=true,
citecolor=darkblue,
linkcolor=reddish,
urlcolor=darkblue,
pdfauthor={},
pdftitle={},
pdfsubject={}
}

\newcommand{\be}{\begin{equation}}
\newcommand{\ee}{\end{equation}}
\newcommand{\bea}{\begin{eqnarray}}
\newcommand{\eea}{\end{eqnarray}}
\newcommand{\beas}{\begin{eqnarray*}}
\newcommand{\eeas}{\end{eqnarray*}}

\def\({\left(}
\def\){\right)}

\newcommand{\rd}{{\rm d}}
\newcommand{\vp}{\varphi}

\def\gsim{ \lower .75ex \hbox{$\sim$} \llap{\raise .27ex \hbox{$>$}} }
\def\lsim{ \lower .75ex \hbox{$\sim$} \llap{\raise .27ex \hbox{$<$}} }

\usepackage{xcolor,colortbl}

\begin{document}
\def\thefootnote{\fnsymbol{footnote}}

~\vspace{1cm}

\begin{center}
\huge{\textbf{Multiple soft limits of cosmological correlation functions}} \\[0.5cm]
\vspace{.2cm}
 
\large{Austin Joyce,${}^{\rm a}$ Justin Khoury${}^{\rm b}$ and Marko Simonovi\'c${}^{\rm c,d}$}
\\[0.5cm]

\small{
\textit{$^{\rm a}$ Enrico Fermi Institute and Kavli Institute for Cosmological Physics,\\ University of Chicago, Chicago, IL 60637}}

\vspace{.2cm}

\small{
\textit{$^{\rm b}$ Center for Particle Cosmology, Department of Physics and Astronomy, \\ University of Pennsylvania, Philadelphia, PA 19104}}

\vspace{.2cm}

\small{
\textit{$^{\rm c}$ SISSA, via Bonomea 265, 34136, Trieste, Italy}}

\vspace{.2cm}

\small{
\textit{$^{\rm d}$ Istituto Nazionale di Fisica Nucleare, Sezione di Trieste, I-34136, Trieste, Italy}}

\vspace{.2cm}

\end{center}

\vspace{.5cm}


\centerline{\small{\bf Abstract}}
\vspace{-0.05cm}
{\small\noindent We derive novel identities satisfied by inflationary correlation functions in the limit where two external momenta are taken to be small. We derive these statements in two ways: using background-wave arguments and as Ward identities following from the fixed-time path integral. Interestingly, these identities allow us to constrain some of the ${\cal O}(q^2)$ components of the soft limit, in contrast to their single-soft analogues. We provide several nontrivial checks of our identities both in the context of resonant non-Gaussianities and in small sound speed models. Additionally, we extend the relation at lowest order in external momenta to arbitrarily many soft legs, and comment on the many-soft extension at higher orders in the soft momentum. Finally, we consider how higher soft limits lead to identities satisfied by correlation functions in large-scale structure.} 
\vspace{0.3cm}
\noindent


\def\thefootnote{\arabic{footnote}}
\setcounter{footnote}{0}

\newpage
\tableofcontents
\newpage

\section{Introduction}
Symmetries play a fundamental role in physics. Classically, they restrict the form of and couplings allowed in the Lagrangian, while quantum-mechanically they constrain correlation functions and S-matrix elements through Ward identities. In cases where symmetries are spontaneously broken, there is still an avatar of the symmetry in the quantum theory; Ward identities tell us about correlation functions in the presence of soft Goldstone modes. 

In the last few years, it has become increasingly clear that this is a useful lens through which to view primordial fluctuations in cosmology. For example, correlation functions of spectator fields on de Sitter space are highly constrained by the linearly realized de Sitter ${\rm SO}(4,1)$ isometries, which at late times act as conformal transformations on spatial slices~\cite{Antoniadis:1996dj,Maldacena:2011nz,Creminelli:2011mw,McFadden:2011kk,Bzowski:2011ab,Kehagias:2012pd,Mata:2012bx}. Perturbations of the inflaton, meanwhile, can be understood as an effective field theory for the Goldstone boson associated with spontaneously broken time translations~\cite{Creminelli:2006xe,Cheung:2007st}.\footnote{This point of view first appeared in the context of the ghost condensate in~\cite{ArkaniHamed:2003uy}.} The same framework has also been applied to the late universe to study systematically dark energy~\cite{Creminelli:2008wc} and modified gravity models~\cite{Gubitosi:2012hu}.

Yet another perspective can be obtained by working in uniform-density gauge, where scalar perturbations are encoded in the well-known variable $\zeta$. In this gauge, $\zeta$ is
immediately recognized as the Goldstone boson (or dilaton) for the spontaneous breaking of the ${\rm SO}(4,1)$ conformal group on ${\mathbb R}^3$, down to its Euclidean subgroup~\cite{Creminelli:2012ed,Hinterbichler:2012nm}:
\be
{\rm SO}(4,1) \longrightarrow {\rm ISO}(3)~.
\label{confbreak}
\ee
Remarkably, this holds for {\it any} spatially-flat Friedmann-Lema\^itre-Robertson-Walker (FLRW) spacetime~\cite{Hinterbichler:2012nm}, except for exact de Sitter space.\footnote{In the de Sitter limit, $\zeta$ decouples,
and the ${\rm SO}(4,1)$ symmetries are restored.} The non-linearly realized symmetries, spatial dilations and special conformal transformations (SCTs), give rise to ``soft pion'' theorems:
\be
\lim_{\vec{q}\to0} \frac{\langle\zeta_{\vec q}\zeta_{\vec{k}_1}\cdots\zeta_{\vec{k}_N}\rangle'}{P(q)} = \left(\delta_{\cal D}+\frac{1}{2} \vec q\cdot \delta_{\vec{\cal K}}\right)\langle\zeta_{\vec{k}_1}\cdots\zeta_{\vec{k}_N}\rangle'\,,
\label{singlesoftrevintro}
\ee
where $\delta_{\cal D}$ and $\delta_{\vec{\cal K}}$ are the dilation and SCT operators in momentum space, and primes denote on-shell correlators. (See main text for details.) This consistency relation has been derived through background-wave arguments~\cite{Maldacena:2002vr,Creminelli:2004yq,Cheung:2007sv,Creminelli:2012ed,Senatore:2012wy,Huang:2006eha}, through Ward identity machinery~\cite{Assassi:2012zq,Hinterbichler:2013dpa,Goldberger:2013rsa,Berezhiani:2013ewa,Pimentel:2013gza}, and holographic arguments~\cite{Bzowski:2012ih,Schalm:2012pi,Ghosh:2014kba}. The power of the consistency relation lies in its generality:~\eqref{singlesoftrevintro} holds in any single-field model, independent of the slow-roll approximation, sound speed of perturbations etc., provided that the background is a dynamical attractor. It holds for a broad class of initial quantum states satisfying certain analyticity conditions~\cite{Berezhiani:2014kga}, which includes the Bunch--Davies state.

Equation~\eqref{singlesoftrevintro} implies that scalar correlation functions in the soft limit $\vec{q}\rightarrow 0$ are fully determined up to ${\cal O}(q^2)$ in terms of a lower-point correlation function. 
The ${\cal O}(q^2)$ piece is physical -- it represents spatial curvature and hence is model-dependent~\cite{Creminelli:2013cga}. By including tensor perturbations $\gamma_{ij}$, it is possible to constrain higher-order
terms in $q$. In fact, there are infinitely-many consistency relations involving a combination of soft scalar and tensor legs, schematically of the form~\cite{Hinterbichler:2013dpa}:
\be
\lim_{\vec{q}\rightarrow 0} {\partial^n \over \partial q^n}
\left( \frac{ \langle \zeta_{\vec q} {\cal O}_{\vec{k}_1,\cdots,\vec{k}_N} \rangle}{P (q)} + 
\frac{ \langle \gamma_{\vec q} {\cal O}_{\vec{k}_1,\cdots,\vec{k}_N} \rangle}{ P_\gamma (q)} \right)
\sim \sum_{a=1}^N {\partial^n \over \partial k^n_a}  \langle {\cal O}_{\vec{k}_1,\cdots,\vec{k}_N}
  \rangle  \, ,
\label{schematicintro}
\ee
where ${\cal O}_{\vec{k}_1,\ldots,\vec{k}_N}$ is an arbitrary operator built out of scalars and tensors. The $n= 0$, $n=1$ relations reproduce~\eqref{singlesoftrevintro} and its soft-tensor analogue. 
The $n\geq 2$ identities only {\it partially} constrain the soft limit of the amplitude at each order. The $3\rightarrow 2$ relations have been checked explicitly up to and including
${\cal O}(q^3)$~\cite{Berezhiani:2014tda}. All of these identities originate from a single, master consistency relation, which follows from the Slavnov--Taylor identity for spatial
diffeomorphisms~\cite{Berezhiani:2013ewa}. See also~\cite{Pimentel:2013gza}. 

The consistency relations~\eqref{schematicintro} are physical statements. They provide stringent {\it null} tests of inflation: observing a violation of any one of these relations would immediately rule
out all standard single-field inflationary models! The presence of entropy perturbations, departures from the attractor solution~\cite{Khoury:2008wj,Cai:2009fn,Chen:2013aj,Dimastrogiovanni:2014ina}, dissipative effects~\cite{LopezNacir:2012rm} and modified initial states~\cite{Holman:2007na,Meerburg:2009ys,Collins:2009pf,Agullo:2010ws,Ashoorioon:2010xg,Dey:2011mj,Chialva:2011hc,Ganc:2011dy,Kundu:2011sg,Dey:2012qp,Agarwal:2012mq,Flauger:2013hra,Aravind:2013lra,Gong:2013yvl,Ashoorioon:2013eia} have all been shown to result in violations of the consistency relations.

In this paper we focus on another probe of the higher-$q$ dependence of correlation functions, namely multiple soft limits. For concreteness, we focus on scalar perturbations only, with symmetry breaking pattern~\eqref{confbreak}.
Our original motivation for this work came from pion physics. Low-energy theorems with two soft pions are sensitive to the non-Abelian nature of the symmetry algebra~\cite{Weinberg:1966kf,ArkaniHamed:2008gz}. For instance, for the symmetry breaking pattern ${\rm SU}(2)_{\rm L}\times {\rm SU}(2)_{\rm R} \longrightarrow {\rm SU}(2)_{\rm D}$, the following holds
\be
\label{nimachiral}
\lim_{\vec q_a,\vec q_b\to0}\langle\pi^a_{\vec q_a}\pi^b_{\vec q_b}\pi^{i_1}_{\vec k_1}\cdots\pi^{i_n}_{\vec k_n} \rangle = \frac{1}{2}\sum_{j}\frac{(\vec q_a-\vec q_b)\cdot \vec k_j}{(\vec q_a+\vec q_b)\cdot \vec k_j}\epsilon^{abc}\langle\pi^{i_1}_{\vec k_1} \cdots X_c\pi^{i_j}_{\vec k_j}\cdots\pi^{i_n}_{\vec k_n}\rangle~,
\ee
where $X_c$ are ${\rm SU}(2)$ generators.  Thus, in the presence of two soft external legs, the remaining hard-momentum modes feel a momentum-dependent group transformation, given by the commutator of the generators associated with the soft modes. The intuition is that when taking two legs to be soft, it matters which momentum is taken to zero first, and this picks out a path through the group. 

In our case, since dilations and SCTs do not commute, one would expect by analogy that the double-soft limit at ${\cal O}(q)$ should depend on the order in which the limit is taken. Here however, the non-commutativity appears in a slightly more subtle form. The point is that we are dealing with space-time symmetries, and as a result only one Goldstone field is needed
to non-linearly realize the conformal group.\footnote{This is a manifestation of what is often called the {\it inverse Higgs effect}~\cite{Ivanov:1975zq}.} Indeed, the fact that the
single-soft relation~\eqref{singlesoftrevintro} constrains correlators at ${\cal O}(q)$ can already be understood as a probe of the de Sitter algebra. Nevertheless, we will see that there is still some trace of non-commutativity in the double-soft relation: in the background wave derivation, it turns out that in order to induce a physical long-wavelength mode at  ${\cal O}(q)$ by a coordinate transformation, we must do a dilation followed by a SCT {\it in that order}. If we choose to do the transformations in the opposite order, we must do an additional compensating SCT to match to a physical mode, capturing the non-commutativity of the de Sitter group.

The results we derive offer a new set of consistency relations for inflationary correlators. For most of the paper we will focus on two soft legs. After briefly reviewing the conformal symmetries of adiabatic
modes (Sec.~\ref{infsymm}), we turn to the derivation of double-soft consistency relations, following two different methods: $i)$ through the background wave method (Sec.~\ref{backgroundwave}); $ii)$ by applying the Ward identity machinery to the fixed-time 1PI action~\cite{Goldberger:2013rsa,Berezhiani:2013ewa} (Sec.~\ref{wardidentity}). These two approaches each have advantages and limitations. The higher-soft limits are most easily phrased in terms of 1PI vertices, but the resummation from vertices to correlation functions becomes increasingly tedious for higher-point correlation functions. Multiple-soft limits are non-trivial to derive in the background wave argument, but the derivation holds for arbitrary number
of hard legs. 

Our results can be summarized as follows: at zeroth order in the soft momenta, the double-soft consistency relation following from dilation invariance is given by
\be
\lim_{\vec q_1, \vec q_2 \to 0}\frac{\langle\zeta_{\vec q_1}\zeta_{\vec q_2}\zeta_{\vec k_1}\cdots\zeta_{\vec k_N}\rangle'}{P(q_1)P (q_2)} = \frac{\langle\zeta_{\vec q_1}\zeta_{\vec q_2}\zeta_{-\vec q}\rangle'}{P(q_1)P (q_2)}\delta_{\cal D} \langle\zeta_{\vec k_1}\cdots\zeta_{\vec k_N}\rangle' +  \delta_{\cal D}^2 \langle\zeta_{\vec k_1}\cdots\zeta_{\vec k_N}\rangle'~.
\label{bwave2dilinrto}
\ee
The first term on the right-hand side comes from the exchange diagram in which two soft modes combine and are attached to the hard modes via a soft internal line (left in Fig.~\ref{2dilationfig}). The second term 
comes from the diagram where both soft lines come from a single vertex (right in Fig.~\ref{2dilationfig}). Similar relations have appeared in the literature before~\cite{Huang:2006eha,Senatore:2012wy},
however these earlier results missed the second term. For resonant non-Gaussianities, for instance, the second term dominates in the double-squeezed limit.

At higher-order in the soft momenta, the double-soft relation takes the following form, which generalizes~\eqref{singlesoftrevintro} to the double-soft case:
\begin{align}
\nonumber
& \lim_{\vec q_1,\vec q_2\to0}\frac{\langle\zeta_{\vec q_1}\zeta_{\vec q_2}\zeta_{\vec k_1}\cdots\zeta_{\vec k_N}\rangle'}{P(q_1)P(q_2)} = \\\label{2sctsintro}
& ~\frac{\langle\zeta_{\vec q_1}\zeta_{\vec q_2}\zeta_{-\vec q}\rangle'}{P(q_1)P(q_2)}\left[ \left(\delta_{\cal D}+\frac{1}{2}\vec q\cdot\delta_{\cal\vec K}\right)\langle\zeta_{\vec k_1}\cdots\zeta_{\vec k_N}\rangle' +q_iq_j \frac{\partial}{\partial \kappa_{ij}} \langle\zeta_{\vec k_1}\cdots\zeta_{\vec k_N}\rangle'_\kappa \right] \\
\nonumber
& ~ +\left(\delta_{\cal D}^2+\frac{1}{2}\vec q\cdot\delta_{\cal \vec K}\delta_{\cal D}+\frac{1}{4}q^i_1 q^j_2\delta_{{\cal K}^i}\delta_{{\cal K}^j}\right)\langle\zeta_{\vec k_1}\cdots\zeta_{\vec k_N}\rangle' + \frac{q_{1i}q_{2j}}{2}\left(\delta^{ij}\nabla_p^2-2\nabla_p^i\nabla_p^j\right)\frac{\langle\zeta_{\vec p}\zeta_{\vec k_1}\cdots\zeta_{\vec k_N}\rangle'}{P(p)}\bigg\rvert_{\vec p\to0} ~,
\end{align}
where $\vec q=\vec q_1+\vec q_2$. The last term in the second line, which can equivalently by re-written as~\cite{Creminelli:2013cga}
\be
\label{curvconsis}
q_iq_j \frac{\partial}{\partial \kappa_{ij}} \langle\zeta_{\vec k_1}\cdots\zeta_{\vec k_N}\rangle'_\kappa = \frac{q_iq_j}{2}\nabla_p^i\nabla_p^j\frac{\langle\zeta_{\vec p}\zeta_{\vec k_1}\cdots\zeta_{\vec k_N}\rangle'}{P(p)}\bigg\rvert_{\vec p\to0}~,
\ee
captures effect of a single soft internal line produced by the two long modes at order $\mathcal O(q^2)$. At this order, the effect of the soft internal line is equivalent to being in a locally curved anisotropic universe where the curvature $\kappa_{ij}$ is related to the second derivatives of this internal long mode. From this we see that even at order ${\cal O}(q_1\cdot q_2)$ the double squeezed limit contains ``physical" terms that are not related to a change of coordinates. For this reason, at this order, eq.~\eqref{2sctsintro} is not a standard consistency relation. In particular, it relates the soft limit of an $(N+2)$-point function to both the $(N+1)$-point function and the $N$-point function.

In Sec.~\ref{corrcheck}, we perform some explicit checks of these identities in the simplest case of 4-point functions in the double-soft limit. Unfortunately, few examples of 4-point functions have been computed in the literature.
Our checks are therefore limited to resonant non-Gaussianities (Sec.~\ref{1piresonantcheck}) and models with small sound speed (Sec.~\ref{smallcs}). In Sec.~\ref{Nsoft}, we generalize the double-soft results to $N>2$ soft legs.
This is completely straightforward at the level of the 1PI action, though once again the resummation to correlation functions is non-trivial. For simplicity, we will focus on multiple dilations and make an intuitive argument for the form of the identity in terms of correlators. In Sec.~\ref{LSS}, we apply our results to the Large Scale Structure, and generalize known consistency relations~\cite{Creminelli:2013mca, Creminelli:2013poa, Creminelli:2013nua,Valageas:2013cma,Kehagias:2013yd,Peloso:2013zw,Peloso:2013spa,Kehagias:2013rpa,Horn:2014rta} to the double-soft case. Finally, we discuss future research directions in Sec.~\ref{conclu}.

\section{Conformal symmetries and adiabatic modes}
\label{infsymm}

We begin by reviewing the residual symmetries of the gauge-fixed inflationary action, focusing on scalar perturbations. Following~\cite{Hinterbichler:2012nm,Hinterbichler:2013dpa},
we work in $\zeta$-gauge, defined by an unperturbed scalar field
\be
\delta\phi\equiv \phi(\vec{x},t)-\bar{\phi}(t)=0\,,
\ee
and a conformally-flat spatial metric
\be
h_{ij} = a^2(t)e^{2\zeta(\vec{x},t)}\delta_{ij}~.
\label{zetagaugespatial}
\ee
We look for diffeomorphisms that preserve this gauge. To preserve $\delta\phi = 0$, clearly the diffeomorphisms must be purely spatial, $\xi^i(\vec{x}, t)$, but {\it a priori}
can depend on time. To preserve the conformal flatness of $h_{ij}$, the residual spatial diffeomorphisms are just conformal transformations on ${\mathbb R}^3$. This 10-parameter group
includes 3 translations, 3 rotations, 1 dilation and 3 SCTs. We henceforth ignore translations and rotations, since these are linearly realized on $\zeta$
and therefore do not give rise to soft-pion theorems. 

Dilations and SCTs act on the coordinates as (see, {\it e.g.},~\cite{DiFrancesco:1997nk})
\bea
\nonumber
 x^i &\longmapsto & e^{\lambda(t)} x^i~,~~~~~~~~~~~~~~~~~~~~~~~~~~~~({\rm dilation})\\
 x^i &\longmapsto & \frac{x^i - b^i(t) \vec x^2}{1-2  \vec x\cdot \vec b(t) +\vec x^2 \vec b^2(t) }~.~~~~~~~~~~~~({\rm SCT})
\eea
Under these transformations, the spatial part of the line element transforms as
\be
\delta_{ij}\rd x^i\rd x^j \longmapsto  \Omega^2(\vec{x},t)\delta_{ij}\rd x^i\rd x^j~,
\ee
where the conformal pre-factor is given by
\bea
\nonumber
\Omega_{\rm dil}(\vec{x},t) &=& e^{\lambda(t)}~,\\
\Omega_{\rm SCT}(\vec{x},t) &=& \Big( 1-2 \vec x\cdot \vec b(t) + \vec x^2 \vec b^2(t) \Big)^{-1} ~.
\label{omegas}
\eea
At the infinitesimal level, the Killing vectors that generate these transformations are
\bea
\nonumber
\xi^i_{\rm dil} &=& \lambda(t) x^i\,;\\
\xi^i_{\rm SCT} &=& 2\vec x\cdot \vec b(t) x^i-\vec x^2 b^i(t)\,. 
\eea
The change in the spatial line element by this conformal factor can be absorbed in a redefinition of $\zeta$; which infinitesimally transforms as
\bea
\nonumber
\delta_{\rm dil}\zeta &=& \lambda(t)\left( 1+\vec x\cdot\vec\partial\zeta\right)~, \\
\delta_{\rm SCT} \zeta &=& b_i(t)\bigg(2x^i+\left(2x^ix^j\partial_j-\vec x^2\partial^i\right)\zeta\bigg)~.
\label{zetadilsct}
\eea

Thus far these transformations are just diffeomorphisms, and as such map solutions of the equations of motion to other solutions.
However, the $\delta\zeta$ profiles induced by~\eqref{zetadilsct} do not preserve boundary conditions: they map field configurations
which fall off at infinity into those which do not. If we want~\eqref{zetadilsct} to represent the long-wavelength limit of a {\it physical mode} --
with suitable fall-off behavior at infinity -- then we must check whether the profiles thus induced can satisfy the constraint equations
away from $\vec{k} = 0$. In other words, these profiles cannot ``accidentally'' solve the equations simply because they are being hit by
spatial derivatives. 

This was checked carefully in~\cite{Creminelli:2012ed,Hinterbichler:2012nm,Hinterbichler:2013dpa} by generalizing Weinberg's
original argument~\cite{Weinberg:2003sw}. The result is that only a subset of the transformations can be extended to a
physical mode. For starters, the parameters of the transformations must be time-independent: $\dot{\lambda} = 0$, $\dot{b}^i = 0$.
Moreover, a SCT must be accompanied by a time-dependent translation~\cite{Creminelli:2012ed,Hinterbichler:2012nm}:
\be
\xi^i =\xi^i_{\rm SCT} - 2b^i \int^t \frac{{\rm d}t'}{H(t')}\,.
\label{SCTadia}
\ee
These are the adiabatic modes of scalar perturbations: field profiles that can be induced by a coordinate transformation which are the $q\to0$ limit of a physical field configuration. In~\cite{Hinterbichler:2013dpa,Berezhiani:2014tda}, the analysis was generalized to include tensors. In this case,
one finds an infinite number of residual global diffeomorphisms which are non-linearly realized on the perturbations. 

Equation~\eqref{SCTadia} can be understood physically by considering the linear solution with a long-wavelength mode $\zeta_{\rm L}$. After solving the constraint equations for the lapse function and shift vector,
the perturbed line element to leading order in gradients~\cite{Maldacena:2002vr, Boubekeur:2008kn} takes the form (see~\cite{Weinberg:2003sw,Creminelli:2012ed,Hinterbichler:2012nm,Hinterbichler:2013dpa} for details)
\be
\rd s^2 = -\rd t^2 - \frac 2 H \partial_i \zeta_{\rm L} \rd x^i \rd t + a^2(t) \left(1+2\zeta_{\rm L}\right) \delta_{ij}\rd x^i\rd x^j~,
\label{zetagaugelinelement}
\ee
where we have specialized to single-field, slow-roll inflation for simplicity. The aim is to show that this metric can be generated by performing a suitable change of coordinates on an unperturbed, homogeneous solution.
Alternatively, we can think of removing a long-wavelength $\zeta$ mode of this form by performing the inverse transformation.
To generate the desired $h_{ij}$ while remaining in $\zeta$-gauge, as discussed earlier, the only allowed transformations are spatial dilations and SCTs. These preserve the conformal flatness of $h_{ij}$. 
Since $\dot{\zeta} \sim k^2\zeta$, $\zeta_{\rm L}$ is constant in time at linear order in gradients, hence the conformal transformations must be time-independent as well.
Working at linear order in $\zeta_{\rm L}$, it is easy to check that the transformation 
\be
\label{linear_dil_sct}
x^i \longmapsto x^i +  \lambda x^i + 2 \vec x\cdot \vec b\, x^i - \vec x^2 b^i~,
\ee
generates the desired spatial metric provided we make the identification
\be
\zeta_{\rm L} = \lambda + 2 \vec b \cdot \vec x\,.
\label{longmode}
\ee

In order to generate the complete metric~\eqref{zetagaugelinelement}, we need an additional change of coordinates to induce the shift vector component $g_{0i} = - H^{-1} \partial_i \zeta_{\rm L}$.
This is achieved by performing a time-dependent translation
\be
x^i \longmapsto x^i - \frac{1}{2H^2 a^2}\partial^i \zeta_{\rm L} \;.
\label{timedeptrans}
\ee
This confirms the result~\eqref{SCTadia} that the linear-gradient adiabatic mode corresponds to a time-independent SCT combined with a particular time-dependent translation. 
Since all correlation functions we are interested in are translationally invariant, in practice the additional transformation~\eqref{timedeptrans} will be of no consequence to our discussion 
and will henceforth be ignored. Although we have focused here on slow-roll inflation for concreteness, it can be shown that the construction of the adiabatic long mode is similar in any model of single-field inflation~\cite{Creminelli:2012ed}. Furthermore, short-wavelength modes can be included non-perturbatively. The relevant part of the coordinate transformation is always related to dilations and SCTs. 

\section{Background-wave derivation}
\label{backgroundwave}
One way to derive the consistency relations is based on the observation that in any single-field model, a physical long mode is indistinguishable from a coordinate transformation. This has been used to derive the inflationary consistency relations~\cite{Creminelli:2004yq,Cheung:2007sv, Creminelli:2012ed, Senatore:2012wy, Huang:2006eha}, consistency relations for the Conformal Mechanism~\cite{Creminelli:2012qr} and consistency relations for Large Scale Structure in the late universe~\cite{Creminelli:2013mca, Creminelli:2013poa, Creminelli:2013nua,Valageas:2013cma,Peloso:2013zw,Kehagias:2013yd,Peloso:2013spa,Kehagias:2013rpa,Horn:2014rta}. The core of the derivation is based on the construction of a long adiabatic mode~\cite{Weinberg:2003sw}. In this Section, we are going to use this technique to derive the consistency relations for the multiple soft limits of inflationary correlation functions.

\subsection{Single-soft relation}
\label{bwaveintro}

Once the long mode has been identified with a change of coordinates, as in~\eqref{longmode}, the derivation of the consistency relation is straightforward. By performing a change of coordinates, we can relate two different solutions, one with and one without a long mode. Consequently, correlation functions in the presence of a long mode must equal correlation functions without the long mode but in transformed coordinates: 
\be
\langle\zeta(\vec x_1)\cdots\zeta(\vec x_N)\rangle_{\zeta_{\rm L}} = \langle\zeta(\vec{\tilde{x}}_1)\cdots\zeta(\vec{\tilde{x}}_N)\rangle \,,
\label{consisbwave}
\ee
where $\tilde{x}^i = x^i +  \zeta_{\rm L}^0 x^i + \vec x\cdot \vec{\partial} \zeta_{\rm L}^0\, x^i - \frac{1}{2} \vec x^2\partial^i\zeta_{\rm L}^0$ and $\zeta_L^0$ is the value of the long-wavelength around the (arbitrarily chosen) origin of coordinates. Expanding the right-hand side of \eqref{consisbwave} to linear order in $\zeta_{\rm L}$,
we obtain
\bea
\nonumber
\langle\zeta(\vec x_1)\cdots\zeta(\vec x_N)\rangle_{\zeta_{\rm L}} &\simeq & \langle\zeta(\vec x_1)\cdots\zeta(\vec x_N)\rangle \\
& & + \sum_{a=1}^N \bigg(\zeta_{\rm L}^0 x_a^i - \frac{1}{2} \partial_j \zeta_{\rm L}^0 \left(\vec{x}_a^2 \delta^{ij} - 2 x_a^i x_a^j\right)\bigg) \frac{\partial}{\partial x_a^i} \langle\zeta(\vec x_1)\cdots\zeta(\vec x_N)\rangle \,.
\eea
We then average both sides over the long mode. Using the definition of a conditional probability, the left-hand side becomes an $(N+1)$-point function: 
\be
\langle \zeta_{\rm L} \langle\zeta(\vec x_1)\cdots\zeta(\vec x_N)\rangle_{\zeta_{\rm L}} \rangle = \langle \zeta_{\rm L}\rangle \langle\zeta(\vec x_1)\cdots\zeta(\vec x_N)\rangle_{\zeta_{\rm L}}=
\langle \zeta_{\rm L} \zeta(\vec x_1)\cdots\zeta(\vec x_N)\rangle,
\ee
On the right-hand side, meanwhile, the first term gives a contribution proportional to $\langle \zeta_{\rm L} \rangle$, which vanishes.
The remaining terms are proportional to the 2-point function of $\zeta_{\rm L}$ times an operator acting on the short modes. Transforming 
to Fourier space and removing delta functions, the result is the standard consistency condition relating $(N+1)$ and $N$-point functions~\cite{Creminelli:2012ed}:
\be
\lim_{\vec{q}\to0} \frac{\langle\zeta_{\vec q}\zeta_{\vec k_1}\cdots\zeta_{\vec k_N}\rangle'}{P(q)} = \left(\delta_{\cal D}+\frac{1}{2} \vec q\cdot \delta_{\vec{\cal K}}\right)\langle\zeta_{\vec k_1}\cdots\zeta_{\vec k_N}\rangle'\,,
\label{singlesoftrev}
\ee
where the primed correlators $\langle\cdots\rangle'$ are correlators without the momentum conserving delta function:
\be
\langle {\cal O}_{\vec{k}_1,\ldots,\vec{k}_N}\rangle = (2\pi)^3 \delta^3(\vec{k}_1 + \ldots + \vec{k}_N)\langle {\cal O}_{\vec{k}_1,\ldots,\vec{k}_N}\rangle'\,. 
\ee
The Fourier-space dilation and SCT operators are defined by
\begin{align}
\nonumber
\delta_{\cal D} &= -3(N-1) - \sum_{a=1}^N \vec k_a \cdot \vec\nabla_{k_a}\,;\\
\delta_{{\cal K}^i} &= \sum_{a=1}^N\left(-6\nabla_{k_a}^i+ k_a^i\nabla_{k_a}^2-2\vec k_a\cdot\vec\nabla_{k_a}\nabla_{k_a}^i\right)~,
\label{fourierops}
\end{align}
and $\nabla_{k_a}^i \equiv \frac{\partial}{\partial k_a^i}$. Thus $(N+1)$-point correlation functions with a soft mode are fully determined up to ${\cal O}(q^2)$ in terms of $N$-point functions without the soft mode.

\subsection{Double-soft dilation result}
\label{doublesoftdil}
In order to generalize the preceding argument to multiple soft legs, we must find a change of coordinates that induces a non-linear background of the long mode. Neither the change of coordinates nor the metric can be linearized.
For simplicity let us start by considering two soft modes, which requires working at quadratic order in $\zeta_{\rm L}$. As we will see, this example is sufficiently rich to capture all
the relevant issues in the problem. We will discuss how our results generalize to additional soft modes at the end of the paper.

We first focus on the relation at zeroth order in the soft momentum, corresponding to dilations only. The only non-trivial part of the metric is the spatial line element:
\be
\rd \ell^2 = a^2(t) e^{2\zeta_{\rm L}} \delta_{ij} \rd x^i \rd x^j \;,
\label{spatiallineelement}
\ee
where $\zeta_{\rm L}$ is constant in both space and time. Clearly, $\zeta_{\rm L}$ is generated non-linearly by a dilation of the spatial coordinates
\be
x^i \  \longmapsto \tilde x^i = e^{\zeta_{\rm L}} x^i  \;.
\ee
Therefore, evaluating an $N$-point function in the presence of a long-wavelength mode is equivalent to working in rescaled coordinates:
\be
\langle\zeta(\vec x_1)\cdots\zeta(\vec x_N)\rangle_{\zeta_{\rm L}} =\langle\zeta(\vec{\tilde{x}}_1)\cdots\zeta(\vec{\tilde{x}}_N)\rangle \,.
\label{2softdil1}
\ee
To derive the double-soft relation, we must expand the change of coordinates up to second order in the long mode:
\be
\delta x^i = \tilde{x}^i - x^i \simeq  \left(\zeta_{\rm L}+\frac{1}{2}\zeta_{\rm L}^2\right) x^i\,.
\ee
Expanding the right-hand side of~\eqref{2softdil1} to quadratic order in $\zeta_{\rm L}$, we obtain
\be
\label{split_variation}
\langle\zeta(\vec x_1)\cdots\zeta(\vec x_N)\rangle_{\zeta_{\rm L}} =\langle\zeta_1\cdots\zeta_N\rangle+\delta\langle\zeta_1\cdots \zeta_N\rangle \;,
\ee
where
\begin{align}
\label{dil_expansion}
\nonumber
\delta\langle\zeta_1\cdots\zeta_N\rangle &=  \sum_{a=1}^N\delta \vec x_a\cdot\vec\nabla_a \langle\zeta_1\cdots\zeta_N\rangle+\frac{1}{2}\sum_{a,b = 1}^N \delta x_a^i\delta x_b^j \nabla^i_a\nabla_b^j\langle\zeta_1\cdots\zeta_N\rangle \\
& =\zeta_{\rm L}\sum_{a=1}^N\vec x_a\cdot\vec\nabla_a \langle\zeta_1\cdots\zeta_N\rangle+\frac{\zeta_{\rm L}^2}{2}\left(\sum_{a=1}^N\vec x_a\cdot\vec\nabla_a+\sum_{a,b = 1}^N x_a^i x_b^j \nabla^i_a\nabla_b^j\right)\langle\zeta_1\cdots\zeta_N\rangle \;.
\end{align}
We will eventually average this relation over two long modes $\zeta_{\rm L}(\vec x) \zeta_{\rm L}(\vec y)$. Before doing so, note that the first term in~\eqref{split_variation}, once averaged over the long modes, will contribute to disconnected correlation functions. We will therefore drop it from our analysis by considering only connected correlators. Averaging the second term will give two kind of contributions -- one that is proportional to the 3-point function of the long modes and a second one that is proportional to two power spectra of the long modes. These two terms will come from contributions in the second line of~\eqref{dil_expansion} respectively proportional to $\zeta_{\rm L}$ and $\zeta_{\rm L}^2$.
 
Rewriting~\eqref{dil_expansion} in Fourier space, we get
\begin{align}
\delta\langle\zeta_1\cdots\zeta_N\rangle &= \lim_{\vec q\to0}\int\frac{\rd^3q}{(2\pi)^3}\frac{\rd^3k_1}{(2\pi)^3}\cdots\frac{\rd^3k_N}{(2\pi)^3}e^{i \vec q\cdot \vec y} (2\pi)^3\delta^{(3)}(\vec P)\langle\zeta_{\vec k_1}\cdots\zeta_{\vec k_N}\rangle' \\\nonumber
&\times\left[\zeta_{\vec q} \sum_{a=1}^N\vec k_a\cdot\vec\nabla_{k_a} +\frac{1}{2}\int\rd^3 Q\,\zeta_{\vec q}\zeta_{\vec Q-\vec q}\left(\sum_{a=1}^N\vec k_a\cdot\vec\nabla_{k_a}+\sum_{a,b=1}^N k_a^i k_b^j \nabla_{k_a}^i\nabla_{k_b}^j\right)\right]e^{i\sum_a\vec k_a\cdot \vec x_a} \;,
\end{align}
where $\vec P = \vec k_1 + \cdots + \vec k_N$. We would like to rewrite this expression in a form where all $k$-derivatives act on the primed correlation function $\langle\zeta_{k_1}\cdots\zeta_{k_N}\rangle'$. Doing this requires a fair amount of integration by parts, which will result in derivatives acting on the delta function $\delta(\vec P)$. These terms come in two flavors:
\bea
\nonumber
f(\vec k_a) \sum_{a=1}^N k_{ai} \nabla_{k_{ai}} \delta(\vec P) e^{i\sum_a \vec k_a\cdot \vec x_a}  &=& f(\vec k_a) P_i \nabla_{P_i} \delta(\vec P) e^{i\sum_a\vec k_a\cdot \vec x_a}\,;\\
g(\vec k_a) \sum_{a,b=1}^N k_{ai}k_{bj} \nabla_{k_{ai}} \nabla_{k_{bj}} \delta(\vec P) e^{i\sum_a\vec k_a\cdot \vec x_a} &=& g(\vec k_a) P_iP_j \nabla_{P_i} \nabla_{P_j} \delta(\vec P) e^{i\sum_a\vec k_a\cdot \vec x_a} \;.
\eea
Integrating by parts again, the only non-zero terms are those where the derivatives act on $P_i$, since all others are proportional to the integral of $P_i\delta(\vec P)$ which gives zero. The result is
\bea
\nonumber
f(\vec k_a) \sum_{a=1}^N k_{ai} \nabla_{k_{ai}} \delta(\vec P) e^{i\sum_a\vec k_a\cdot \vec x_a} &=& - 3f(\vec k_a) \delta(\vec P) e^{i\sum_a\vec k_a\cdot \vec x_a}\,;  \\
g(\vec k_a) \sum_{a=1}^N \sum_{b=1}^N k_{ai}k_{bj} \nabla_{k_{ai}} \nabla_{k_{bj}} \delta(\vec P) e^{i\sum_a\vec k_a\cdot \vec x_a} &=& 12g(\vec k_a) \delta(\vec P) e^{i\sum_a\vec k_a\cdot \vec x_a} \;.
\eea
Putting all contributions together, the variation of the $N$-point function takes the form
\be
\delta\langle\zeta_{\vec k_1}\cdots\zeta_{\vec k_N}\rangle' = \lim_{\vec q\to0}\left(\zeta_{\vec q}\, \delta_{\cal D} +\frac{1}{2}\int\rd^3 Q\,\zeta_{\vec q}\zeta_{\vec Q-\vec q}\, \delta_{\cal D}^2\right)\langle\zeta_{\vec k_1}\cdots\zeta_{\vec k_N}\rangle'~,
\ee

\begin{figure}
\centering
\begin{subfigure}{}
\includegraphics[width=2.5in]{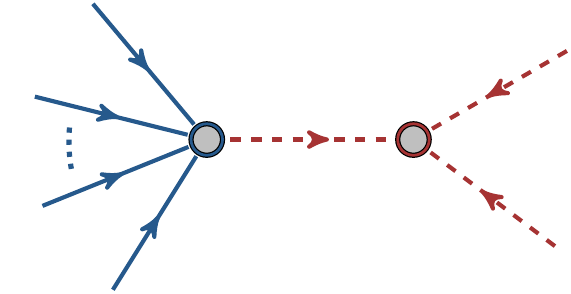}
\end{subfigure}
~~~~~~~~~~
\begin{subfigure}{}
\includegraphics[width=1.9in]{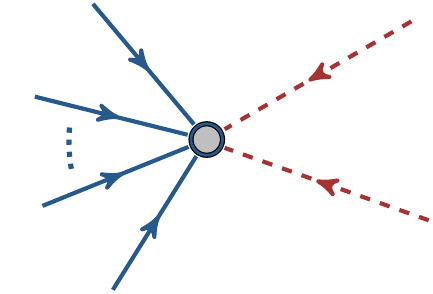}
\end{subfigure}
\caption{\label{2dilationfig}{\small Contributions to the double squeezed limit~\eqref{bwave2dil}. {\it Left}: Two soft modes combine and are connected to the hard modes via a soft internal line. {\it Right}: Both soft lines come from the same vertex as the hard modes, resulting in a double-dilation on the remaining hard modes.}}
\end{figure}

The final step is to, insert this into~\eqref{split_variation}, multiply the resulting expression by two long-wavelength modes $\zeta_{\vec q_1}$ and $\zeta_{\vec q_2}$ and average. In this way we obtain the {\it double-soft relation for dilations}:
\be
\lim_{\vec q_1, \vec q_2 \to 0}\frac{\langle\zeta_{\vec q_1}\zeta_{\vec q_2}\zeta_{\vec k_1}\cdots\zeta_{\vec k_N}\rangle'}{P(q_1)P (q_2)} = \frac{\langle\zeta_{\vec q_1}\zeta_{\vec q_2}\zeta_{-\vec q}\rangle'}{P(q_1)P (q_2)}\delta_{\cal D} \langle\zeta_{\vec k_1}\cdots\zeta_{\vec k_N}\rangle' +  \delta_{\cal D}^2 \langle\zeta_{\vec k_1}\cdots\zeta_{\vec k_N}\rangle'~.
\label{bwave2dil}
\ee
This formula can be understood quite simply from a diagrammatic perspective, as shown in Fig.~\ref{2dilationfig}. The two contributions to the soft limit come from different ways that an $N$-point function can factorize when two of the external lines are taken to be soft. The first term in~\eqref{bwave2dil} describes two soft modes combining and linking to the hard modes via a soft internal line (left in Fig.~\ref{2dilationfig}). This causes the diagram to factorize into a $3$-point function with no hierarchy between the modes times a two point function rescaled by the presence of a long-wavelength mode. The second term in~\eqref{bwave2dil} corresponds to the case where both soft lines come from a single vertex, which induces a double dilation on the remaining hard modes.

Equation~\eqref{bwave2dil} is a generalization of Maldacena's consistency relation for a single soft external leg. Similar relations have appeared in the literature before~\cite{Huang:2006eha,Senatore:2012wy}. However, these relations {\it only} contained the first term on the right-hand side of~\eqref{bwave2dil}. The second term in our result comes from expanding everything up to second order in $\zeta_{\rm L}$ consistently. It is important to stress that in some cases these two terms may be of different order in slow-roll parameters, with one being suppressed compared to the other. In single-field models with reduced speed of sound, for instance, the 3-point function is not slow-roll suppressed, while the second term is higher-order in slow-roll parameters. However, this is not generically true. For example, in standard slow-roll inflation both terms are expected to be equally relevant (second order in slow-roll parameters). Further -- as we are going to show later in explicit checks -- for the model of resonant non-Gaussianities the second term is dominant in the double-squeezed limit of the correlation functions.

Additionally, the presence of the second term in~\eqref{bwave2dil} is essential to recover the correct result in the hierarchical limit $q_1\ll q_2$.
In this limit we can apply Maldacena's consistency relation (equation~\eqref{singlesoftrev} at lowest order in $q$) to the 3-point function $\langle\zeta_{\vec q_1}\zeta_{\vec q_2}\zeta_{-\vec q}\rangle'$:
\bea
\nonumber
\lim_{\vec q_1\ll \vec q_2\to0}\frac{\langle\zeta_{\vec q_1}\zeta_{\vec q_2}\zeta_{\vec k_1}\cdots\zeta_{\vec k_N}\rangle' }{P(q_1)}&=& \delta_{\cal D}P(q_2)~\delta_{\cal D}\langle\zeta_{\vec k_1}\cdots\zeta_{\vec k_N}\rangle'+P(q_2)\delta_{\cal D}^2\langle\zeta_{\vec k_1}\cdots\zeta_{\vec k_N}\rangle'\\
&=& \delta_{\cal D}\Big(P(q_2)\delta_{\cal D}\langle\zeta_{\vec k_1}\cdots\zeta_{\vec k_N}\rangle'\Big)~.
\eea
This is precisely the answer one would obtain by applying Maldacena's relation twice on an $(N+2)$-point function. 

\subsection{First order in derivatives}
\label{dilsctsec}

The above derivation is valid to zeroth order in the gradient expansion. In this case, as we have seen, the long mode is a constant background that can be removed non-linearly
by performing a dilation of the spatial coordinates. We would like to extend the result to higher-order in the gradient expansion and allow the long modes to have homogeneous
gradients. For simplicity, we focus again on the case where we have two soft modes. This means that we must consistently work at second order in $\zeta_{\rm L}$. 
As a first step, for the time being we will work to first order in the gradient expansion. In other words, one of the long modes is a linear gradient profile, while the
other is constant. The general, second-order calculation where both soft modes have a linear gradient profile will be treated in Sec.~\ref{fullcoord}.

Up to second order in $\zeta_{\rm L}$ and first order in the gradient expansion, the spatial line element is given by 
\bea
\label{spatial_sct_d}
\nonumber
\rd \ell^2 &=& a^2(t) e^{2\zeta_{\rm L}} \rd \vec x^2  \\
&\simeq & a^2(t) \bigg( 1+2\zeta_{\rm L}^{0} +2(\zeta_{\rm L}^{0})^2+ 2\vec x\cdot\vec\nabla \zeta_{\rm L}^0 + 4 \zeta_{\rm L}^0 \vec x\cdot\vec\nabla \zeta_{\rm L}^0 \bigg)\rd \vec x^2~,
\eea
where we have expanded the long mode around the origin 
\be
\zeta_{\rm L}(x) = \zeta_{\rm L}^0 + \vec x\cdot\vec\nabla \zeta_{\rm L}^0+\cdots~.
\ee
The superscript indicates that the function is evaluated at the origin. Aside from the spatial metric, the $g_{0i}$ components are also non-zero at linear order in gradients -- see~\eqref{zetagaugelinelement}. Dilations do not change this part of the metric even non-linearly and, as in the single-soft case, we expect that at this order in derivatives $g_{0i}$ can be removed by a time-dependent spatial translation. Translations are linearly realized on $\zeta$ and hence do not affect the correlation functions. We therefore ignore this transformation and focus on the spatial metric~\eqref{spatial_sct_d}.  

Obviously, the change of coordinates corresponding to the adiabatic long mode at this order will involve one dilation and one SCT. However, given that these two transformations do not commute, it is not immediately clear in which order they should be performed. The choice is, of course, not arbitrary; only one of these orderings will generate a physical solution with metric~\eqref{spatial_sct_d}. Using~\eqref{omegas}, we see that doing a SCT first followed by a dilation gives\footnote{Here we work to linear order in $\vec b$, but keep terms of order $\lambda \vec b$ and $\lambda^2$. This is because we will see that $\vec b \sim \vec\nabla\zeta$, so we are working to first order in gradients, but second order in $\zeta$.}
\be
(\mathrm D\circ \mathrm{SCT}) \rd \vec x^2 = \( 1+2\lambda +2\lambda^2 + 4\vec b \cdot \vec x + 12 \lambda \vec b \cdot \vec x + \cdots \) \rd \vec x^2  \;,
\label{scththendil}
\ee 
while the other way around gives
\be
(\mathrm {SCT} \circ \mathrm D) \rd \vec x^2 = \( 1+2\lambda +2\lambda^2 + 4\vec b \cdot \vec x + 8 \lambda \vec b \cdot \vec x + \cdots \) \rd \vec x^2  \;.
\label{dilthensct}
\ee 
Comparing with \eqref{spatial_sct_d}, we see that only~\eqref{dilthensct} can be matched by choosing $\lambda$ and $\vec b$ suitably:
\be
\lambda = \zeta_{\rm L}^0 \,;\qquad \vec b = \frac{1}{2}\vec \nabla \zeta_{\rm L}^0\,.
\ee
This exactly matches~\eqref{longmode}. In other words, to generate the the line element~\eqref{spatial_sct_d} we must do the dilation {\it first}. We can, of course, choose to do the transformations in the opposite order, but~\eqref{scththendil} differs from a physical long-wavelenth mode by an additional SCT, which must also be tracked. 

The full change of coordinates (up to an irrelevant spatial translation) that is equivalent to a physical long mode up to second order in $\zeta_{\rm L}$ and first order in derivatives is
\be
\delta x^i = \left(\zeta_{\rm L}^0  + \frac{1}{2} (\zeta_{\rm L}^{0})^2 + \left(1 + \zeta_{\rm L}^0\right) \vec x\cdot \vec \nabla \zeta_{\rm L}^0 \right)  x^i - \frac{1}{2} \left(1 + \zeta_{\rm L}^0\right) \vec x^2  \nabla^i \zeta_{\rm L}^0 \;.
\label{coordlingrad}
\ee

Following similar steps as in Sec.~\ref{doublesoftdil}, we can derive the double-soft consistency relation at ${\cal O}(q)$. The starting point is again~\eqref{split_variation} with the variation of an $N$-point function under a change of coordinates
\be
\delta\langle\zeta_1\cdots\zeta_N\rangle =  \sum_{a=1}^N\delta \vec x_a\cdot\vec\nabla_a \langle\zeta_1\cdots\zeta_N\rangle+\frac{1}{2}\sum_{a,b = 1}^N \delta x_a^i\delta x_b^j \nabla^i_a\nabla_b^j\langle\zeta_1\cdots\zeta_N\rangle+\cdots \,,
\label{varNlingrad}
\ee
where $\delta x$ given by~\eqref{coordlingrad}. Terms of order $\lambda^2$ correspond to a double dilation and reproduce the zeroth-order consistency relation~\eqref{bwave2dil}. Terms of order $b^2$ are second-order in derivatives
and can be ignored in our approximation. (We will come include these in Sec.~\ref{fullcoord} below.) Finally, terms of order $b\lambda$ generate new contributions to the consistency relation at linear order in the gradient expansion. Averaging the $N$-point variation~\eqref{varNlingrad} with two long modes and going to momentum space, after a fairly long computation we obtain the double-soft consistency relation at order $q$:
%
\bea
\nonumber
\lim_{\vec q_1, \vec q_2\to0} \frac{\langle\zeta_{\vec q_1}\zeta_{\vec q_2}  \zeta_{\vec k_1}\cdots \zeta_{\vec k_N}\rangle'}{P(q_1)P(q_2)} &=&  \frac{\langle\zeta_{\vec q_1}\zeta_{\vec q_2}\zeta_{-\vec q}\rangle'}{P(q_1)P(q_2)} \delta_{\cal D}\langle\zeta_{\vec k_1}\cdots\zeta_{\vec k_N}\rangle'+  \delta_{\cal D}^2 \langle\zeta_{\vec k_1}\cdots\zeta_{\vec k_N}\rangle'\\
&+&\frac{1}{2} q^i\bigg(\delta_{{\cal K}^i} \delta_{\cal D}\langle\zeta_{\vec k_1}\cdots\zeta_{\vec k_N}\rangle'+ \frac{\langle\zeta_{\vec q_1}\zeta_{\vec q_2}\zeta_{-\vec q}\rangle' }{P(q_1)P(q_2)}\delta_{{\cal K}^i} \langle\zeta_{\vec k_1}\cdots\zeta_{\vec k_N}\rangle' \bigg) \,.
\label{dilthensctconsis}
\eea
The first line matches the double-dilation result~\eqref{bwave2dil}, as we expect. The second line is a linear-gradient correction which arises from doing a dilation followed a by SCT on the hard modes.

Had we chosen to do the SCT first, the operators in the first term of the second line of~\label{dilthensctconsis} would appear in the opposite order, and the compensating SCT required to match to a physical mode would induce an additional term which takes the form of a single ${\delta_{\cal K}^i}$ acting on the $N$-point function. This is equivalent to swapping the ordering of the operators in~\eqref{dilthensctconsis} and using the commutation relation
\be
\left[\delta_{{\cal K}^i}, \delta_{\cal D}\right] = \delta _{{\cal K}^i}~,
\ee
of the conformal algebra.

\subsection{Full coordinate transformation at ${\cal O}(q^2)$}
\label{fullcoord}

To derive consistency relations at the next order in the gradient expansion --  order in the soft momentum -- we must keep second-order terms in derivatives in the expansion~\eqref{spatial_sct_d}. To be consistent, we have to expand the long mode up to this order too. We can write 
\be
\label{zeta_long}
\zeta_{\rm L}(\vec x,t)= \zeta_{\rm L}^0(t) + \vec x\cdot\vec\nabla \zeta_{\rm L}^0 +\frac{1}{2}x^i x^i\nabla_i\nabla_j\zeta_{\rm L}^0 + \cdots\,,
\ee
where we have allowed for time-dependence of the homogeneous part of the long-wavelength mode because this term is of the same order as $\nabla^2\zeta$. This can be explicitly seen by solving the quadratic equation of motion for $\zeta_{\rm L}^0(t)$ to fix (at leading order in slow-roll parameters)~\cite{Creminelli:2013cga}
\be
\zeta_{\rm L}^0(t)=\zeta_{\rm L}^0-\frac{c_s^2}{2 a^2 H^2}(1-2\epsilon) \nabla^2\zeta_{\rm L}^0~.
\ee
With this in hand, we see that to order $\nabla^2\zeta$ the physical long-wavelength $\zeta$ profile is given by
\be
\label{eq:long_phys}
\zeta_{\rm L}(\vec x,t)= \zeta_{\rm L}^0 + \vec x\cdot\vec\nabla \zeta_{\rm L}^0 +\frac{1}{2}x^i x^i\nabla_i\nabla_j\zeta_{\rm L}^0 -\frac{c_s^2}{2 a^2 H^2}(1-2\epsilon) \nabla^2\zeta_{\rm L}^0~.
\ee
It is important to notice that the term $ \sim \nabla_i\nabla_j\zeta$ is related to the curvature of space. Indeed, one can show that the solution in $\zeta$-gauge with the long mode \eqref{eq:long_phys} is equivalent, up to irrelevant change of coordinates, to the metric of a spatially curved universe \cite{Creminelli:2013cga}. The $x$-dependent part of the long mode is related to the spatial curvature while the time dependent homogenous term modifies the scale factor to match the background evolution in a spatially curved universe.

The effects of curvature are of course physical and cannot be removed by a change of coordinates. Therefore, we can conclude that if we want to be completely general and keep all second order terms in derivatives, we cannot write down a consistency relation in the usual sense. The effect of the long mode at this order is unavoidably physical. For example, expanding $a^2(t) e^{2\zeta_{\rm L}}$ to second order, we find a term of the form
\be
\zeta_{\rm L}^0 x^i x^i\nabla_i\nabla_j\zeta_{\rm L}^0 \;.
\ee
Physically, this corresponds to a situation where one long mode in the double squeezed limit is treated at homogeneous level (and it is equivalent to a dilation) while the effect of the other long mode is physical. 

One might still hope that it is possible to write down the consistency relation at least for some of the second order terms. One good guess would be terms proportional to $\vec q_1\cdot\vec q_2$, because we expect them to come from two gradients of the two long modes which are both separately removable by a change of coordinates. However, as we are going to see now, this is not the case, and the physical effects of the long mode at this order are always present.
 
To see this explicitly, let us first focus on the case where the two long modes combine and through a single soft internal line affect the remaining $N$ hard legs. This is one of the two kinds of diagrams that we have as we discussed in previous sections. In this case the amplitude naturally factorizes in the following way
\begin{align}
\label{eq:echange_O(q2)}
& \lim_{\vec q_1,\vec q_2\to0} \langle \zeta_{\vec q_1}\zeta_{\vec q_2}\zeta_{\vec k_1}  \cdots\zeta_{\vec k_N}\rangle' \supset \lim_{\vec q\to0} \frac1{P(q)} \langle\zeta_{\vec q_1}\zeta_{\vec q_2}\zeta_{-\vec q}\rangle' \langle \zeta_{\vec q} \zeta_{\vec k_1}\cdots\zeta_{\vec k_N}\rangle' \nonumber \\
& \quad= \langle\zeta_{\vec q_1}\zeta_{\vec q_2}\zeta_{-\vec q}\rangle' \left(\delta_{\cal D}+\frac{1}{2}\vec q \cdot\delta_{\cal\vec K}\right)\langle\zeta_{\vec k_1}\cdots\zeta_{\vec k_N}\rangle' + \langle\zeta_{\vec q_1}\zeta_{\vec q_2}\zeta_{-\vec q}\rangle' q_iq_j \frac{\partial}{\partial\kappa_{ij}} \langle\zeta_{\vec k_1}\cdots\zeta_{\vec k_N}\rangle'_\kappa \;,
\end{align}
where in the second line we explicitly wrote the effects of the long mode up to $\mathcal O(q^2)$ terms. We get the first contribution using the conformal consistency relation while the second one captures the physical effects of the long mode that we discussed before and that cannot be traded for a change of coordinates. In order to write it we used the ``curvature" consistency relation of \cite{Creminelli:2013cga}
\be
\lim_{\vec q \to0}~\frac{1}{P(q)}\langle \zeta_{\vec q}\zeta_{\vec k_1}  \cdots\zeta_{\vec k_N}\rangle' \supset q_iq_j \frac{\partial}{\partial \kappa_{ij}}\langle\zeta_{\vec k_1}\cdots\zeta_{\vec k_N}\rangle'_{\kappa} \;,
\ee
where the squeezed limit of an $(N+1)$-point function at second order in gradients is proportional to the $N$-point function evaluated in an locally anisotropic curved universe (this is the meaning of the label $\kappa$), with the spatial curvature related to second derivatives of the long mode. This expression can also be equivalently rewritten using~\eqref{curvconsis}. We can explicitly see that the part $\vec q_1\cdot\vec q_2$ in expression \eqref{eq:echange_O(q2)} is related to the curvature of the long mode and not removable by the change of coordinates.

A similar thing happens for the contact diagrams too. For simplicity, in what follows we will ignore the constant part $\zeta_{\rm L}^0$ and focus only on gradients. Expanding the spatial line element to second order, we obtain
\bea
\nonumber
\rd \ell^2 &\sim& \left(1 + 2\zeta_{\rm L} + 2 \zeta_{\rm L}^2\right)\rd\vec x^2\\
&\simeq & \left(1+2\vec x\cdot\vec\nabla \zeta_{\rm L}^0+ 2\left(\vec x\cdot\vec\nabla\zeta_{\rm L}^0\right)^2\right)\rd\vec x^2~.
\label{2ndorderexp1}
\eea
Our intuition tells us that at least part of the long mode at ${\cal O}(q^2)$ should be removable by an SCT at second order
\bea
\nonumber
\rd \ell^2 &\sim& \left(1-2 \vec b\cdot \vec x+\vec b^2 \vec x^2\right)^{-2} \rd\vec x^2  \\
&\simeq& \left(1+ 4\vec b\cdot\vec x +12\left(\vec b\cdot \vec x\right)^2-2\vec b^2\vec x^2\right)\rd\vec x^2~.
\label{2ndorderexp2}
\eea
For the choice $b_i = \frac{1}{2}\nabla_i\zeta_{\rm L}^0$, the linear terms in $\zeta_{\rm L}^0$ match up in the two expressions~\eqref{2ndorderexp1} and~\eqref{2ndorderexp2}. 
However, the quadratic terms have a different structure. In other words, at this order in gradient expansion, preforming a non-linear SCT {\it does not} induce the correct physical long-wavelength mode. The other way to see that the SCT cannot induce a correct profile of the long mode is notice that starting from an unperturbed FRW and doing a SCT we do not generate any spatial curvature. On the other hand, for $h_{ij}= a^2e^{2\zeta_{\rm L}}\delta_{ij}$, the spatial curvature at order we are interested in is proportional to
\be
R^{(3)} \sim -2\nabla_i\zeta_{\rm L}^0 \nabla^i \zeta_{\rm L}^0 \;.
\ee
Therefore, the effect of two gradients of the two long modes is physical, and only partially can be removed by the non-linear SCT. The difference between the physical $\zeta_{\rm L}$ and the profile induced by the SCT in the line element is
\be
\label{difference}
e^{2\vec x\cdot\vec\nabla \zeta_{\rm L}^0} - \Omega_{\rm sct}^2\big\rvert_{b_i =\frac{1}{2}\nabla_i\zeta_{\rm L}^0} = - \left(\vec x\cdot\vec\nabla\zeta_{\rm L}^0\right)^2 + \frac{1}{2}\vec x^2\left(\vec \nabla\zeta_{\rm L}^0\right)^2~.
\ee
The terms on the right-hand side are second order in $\zeta_{\rm L}$ and contain ``curvature" effects. We expect them to contribute to the squeezed limit of an $(N+2)$-point function in the following way
\be
\lim_{\vec q_1,\vec q_2\to0} \langle \zeta_{\vec q_1}\zeta_{\vec q_2}\zeta_{\vec k_1}  \cdots\zeta_{\vec k_N}\rangle' \supset P(q_1)P(q_2) q_{1i} q_{2j} \frac{\partial}{\partial R^{(3)}_{ij}} \langle \zeta_{\vec k_1}\cdots\zeta_{\vec k_N}\rangle'_{\zeta_{\vec q_1}\zeta_{\vec q_2}} \;,
\ee
where $R_{ij}^{(3)}$ is the part of the 3-curvature quadratic in $\zeta$ and the the subscript $\zeta_{\vec q_1}\zeta_{\vec q_2}$ on the correlation function denotes the residual effects of two long modes on the short modes once the SCT is taken into account. We should expect that this $N$-point function in the presence of curvature should be related to an $(N+1)$-point function with a long-wavelength mode $\zeta\sim x^2$. Given the tensor structure of~\eqref{difference}, this curvature contribution will take the form
\be
\label{eq:contact_O(q1q2)}
\lim_{\vec q_1,\vec q_2\to0} \frac{\langle \zeta_{\vec q_1}\zeta_{\vec q_2}\zeta_{\vec k_1}  \cdots\zeta_{\vec k_N}\rangle'}{P(q_1)P(q_2) } \supset q_{1i} q_{2j} \frac{1}{2} \left(\delta^{ij}\nabla_p^2-2\nabla_p^i\nabla_p^j\right)\frac{\langle\zeta_{\vec p}\zeta_{\vec k_1}\cdots\zeta_{\vec k_N}\rangle'}{P(p)}\bigg\rvert_{\vec p\to0} \;.
\ee
We will not explicitly calculate this effect here, but we are going to see later using Ward identities that this is indeed the correct expression.

We are finally ready to write down the consistency relation with two soft modes at order ${\cal O}(q_1\cdot q_2)$. Putting together contributions from the exchange diagrams \eqref{eq:echange_O(q2)}, effects of two long modes shown in eq.~\eqref{eq:contact_O(q1q2)} and the additional terms containing dilations and SCTs at second order, we finally get
\begin{equation}
\boxed{
\begin{aligned}
& \lim_{\vec q_1,\vec q_2\to0}\frac{\langle\zeta_{\vec q_1}\zeta_{\vec q_2}\zeta_{\vec k_1}\cdots\zeta_{\vec k_N}\rangle'}{P(q_1)P(q_2)} = \\\
& ~\frac{\langle\zeta_{\vec q_1}\zeta_{\vec q_2}\zeta_{-\vec q}\rangle'}{P(q_1)P(q_2)}\left[ \left(\delta_{\cal D}+\frac{1}{2}\vec q\cdot\delta_{\cal\vec K}\right)\langle\zeta_{\vec k_1}\cdots\zeta_{\vec k_N}\rangle' +q_iq_j \frac{\partial}{\partial \kappa_{ij}} \langle\zeta_{\vec k_1}\cdots\zeta_{\vec k_N}\rangle'_\kappa \right] \\
& ~ +\left(\delta_{\cal D}^2+\frac{1}{2}\vec q\cdot\delta_{\cal \vec K}\delta_{\cal D}+\frac{1}{4}q^i_1 q^j_2\delta_{{\cal K}^i}\delta_{{\cal K}^j}\right)\langle\zeta_{\vec k_1}\cdots\zeta_{\vec k_N}\rangle' + \frac{q_{1i}q_{2j}}{2}\left(\delta^{ij}\nabla_p^2-2\nabla_p^i\nabla_p^j\right)\frac{\langle\zeta_{\vec p}\zeta_{\vec k_1}\cdots\zeta_{\vec k_N}\rangle'}{P(p)}\bigg\rvert_{\vec p\to0}. \\
\end{aligned}
}
\label{2scts}
\end{equation}
This is the main result of the paper. To summarize, the first line is related to the exchange diagrams in which two long modes combine into a single soft internal line. In the last term in the first line the $N$-point function should be calculated in a locally anisotropic and curved universe (see \cite{Creminelli:2013cga}). The second line follows from performing a dilation and a SCT at second order. The last term is related to the mismatch of the physical line element and the one induced by a SCT at second order in gradients. Notice that in this expression we neglected terms of the order ${\cal O}(q_1^2)$ and ${\cal O}(q_2^2)$, because they are clearly related to the physical effects of the long modes.

One important check of the above identity is that it does indeed reproduce the correct result in the hierarchical limit, where the conformal consistency relation can be applied twice. In the next Section we will offer an alternative derivation of this result using Ward identity machinery.

\section{Ward identity derivation}
\label{wardidentity}

The preceding derivation of double-soft consistency relations was based on the background-wave argument, namely that long-wavelength $\zeta$ modes can be induced or removed by
appropriate symmetry transformations. However, as we saw, this procedure runs into trouble when trying to remove a long-wavelength mode at second-order in gradients. In this Section,
we will show how this deficiency can be circumvented by employing field-theoretic machinery to derive Ward identities associated with the symmetries of Sec.~\ref{infsymm}.

The technology we will employ is the one particle-irreducible (1PI) action Ward identities of~\cite{Goldberger:2013rsa} (see also~\cite{Berezhiani:2013ewa}). These authors 
introduce a $3d$ Euclidean path integral over field configurations at fixed time, which is sufficient to derive equal-time correlation functions. The information about the prior history
(in particular the initial state) is encoded in the wavefunctional.\footnote{See~\cite{Berezhiani:2014kga} for subtleties regarding the initial state and identities it satisfies in this formalism.} In this formalism, one can define a $3d$ vertex functional or 1PI action $\Gamma[\zeta]$ as usual as the 
Legendre transform of the connected generating functional. The 1PI vertices are then given by 
\be
\frac{\delta^{N}\Gamma[\zeta]}{\delta\zeta_{\vec k_1}\cdots\delta\zeta_{\vec k_N}}{\Big \rvert_{\zeta = 0}} \equiv (2\pi)^3\delta^{(3)}(\vec k_1+\cdots+\vec k_N)\Gamma^{(N)}(\vec k_1,\cdots, \vec k_N)~.
\ee
We have factored out the momentum-conserving delta function for convenience, hence the $\Gamma^{(N)}$ are on-shell vertices. In particular, we will always express the
last momentum $\vec k_N$ in terms of the other $(N-1)$ momenta. 

There are two identities satisfied by the $\Gamma^{(N)}$. The first is a consequence of dilation symmetry~\cite{Goldberger:2013rsa}\footnote{The factor of 3 in~\eqref{1PIdilation} comes from commuting the dilation operator past the delta function~\cite{Maldacena:2011nz, Creminelli:2012ed,Goldberger:2013rsa}.} 
\be
\lim_{\vec q\to0}\Gamma^{(N+1)}(\vec q, \vec k_1,\cdots, \vec k_N) = \left(3-{\cal D}_N\right)\Gamma^{(N)}(\vec k_1,\cdots, \vec k_N)~,
\label{1PIdilation}
\ee
where the differential operator ${\cal D}_N$ is defined as
\be
{\cal D}_N = \sum_{a=1}^N \vec k_a\cdot \vec\nabla_{k_a}~,
\ee
The second identity satisfied by $\Gamma^{(N)}$ is a consequence of SCT symmetry:
\be
\lim_{\vec q\to0}\nabla_q^i\Gamma^{(N+1)}(\vec q, \vec k_1,\cdots, \vec k_N) = \frac{1}{2}{\cal S}_N^i\Gamma^{(N)}(\vec k_1,\cdots, \vec k_N)~.
\label{1PIsct}
\ee
where we have defined\footnote{Comparing with~\eqref{fourierops}, we see that $\delta_{{\cal K}^i} =  \sum_{a=1}^N\left(-6\nabla_{k_a}^i+ {\cal S}_a^i \right)$.}
\be
{\cal S}_N^i = \sum_{a=1}^N {\cal S}_a^i~~~~~~~~{\rm with}~~~~~~~ {\cal S}_a^i \equiv  k_a^i\nabla_{k_a}^2-2\vec k_a\cdot\nabla_{k_a}\nabla_{k_a}^i~.
\ee
Since $\Gamma^{(N)}$ is an on-shell vertex, when acting on it the operators ${\cal D}_N$ and ${\cal D}_{N-1}$ are equivalent; similarly for ${\cal S}_N$ and ${\cal S}_{N-1}$.
We will use this fact often when checking the identities. 

We now apply this machinery to derive both the 1PI Ward identities for multiple soft lines, but also to re-derive the statements in terms of correlation functions obtained in Sec.~\ref{backgroundwave}.
It will soon become clear that working in terms of vertices is technically much easier than correlation functions. The two viewpoints are entirely equivalent, of course, and we will show how to derive the correlation function identities~\eqref{bwave2dil},~\eqref{dilthensctconsis} and~\eqref{2scts} starting from the vertex identities.

\subsection{A single soft leg}
\label{1softsec}
The identities~\eqref{1PIdilation} and~\eqref{1PIsct} are statements about the effect of a single soft leg on the remaining hard modes in the 1PI vertices. Although these identities are mathematically true, their relation to observable quantities is somewhat obscure. Indeed, we would like to translate these identities to statements about  correlation functions of $\zeta$. For a single soft leg, this  `resummation' is straightforward~\cite{Goldberger:2013rsa}. Nevertheless, it is worth
briefly reviewing this procedure for the simplest case of the $3$-point function since this will be important in the multiple-soft generalization.

The relation between the 3-point function and vertex is
\be
\langle\zeta_{\vec q}\zeta_{\vec k_1}\zeta_{\vec k_2}\rangle = (2\pi)^3\delta^{(3)}(\vec q+\vec k_1+\vec k_2) P(q)P(k_1)P(k_2) \Gamma^{(3)}(\vec q, \vec k_1, \vec k_2)~,
\ee
where $P(k) = \langle\zeta_{\vec k}\zeta_{-\vec k}\rangle'$ is the power spectrum of the curvature perturbation. Using~\eqref{1PIdilation} and the relations
\be
\Gamma^{(2)}(k) = -\frac{1}{P(k)}~;~~~~~~~~ ~~~~~~~{\cal D}_1\Gamma^{(2)}(k) = -\frac{1}{P^2(k)}(3+{\cal D}_1)P(k)~,
\ee
we can easily deduce Maldacena's consistency relation:
\be
\lim_{q\to0}\frac{\langle\zeta_{\vec q}\zeta_{\vec k_1}\zeta_{\vec k_2}\rangle' }{P(q)}= -\left(3+{\cal D}_1\right) P(k)~.
\label{maldacena3pt}
\ee
Of course this identity relating the 3-point correlation function of $\zeta$ in a particular limit of momentum space to the 2-point function is merely a special case of the more general identity
\be
\lim_{\vec{q}\to0} \frac{\langle\zeta_{\vec q}\zeta_{\vec k_1}\cdots\zeta_{\vec k_N}\rangle'}{P(q)} = -\Big(3(N-1) + {\cal D}_N\Big)\langle\zeta_{\vec k_1}\cdots\zeta_{\vec k_N}\rangle'\,,
\ee
which relates $(N+1)$-point functions to $N$-point ones.

The conformal vertex identity~\eqref{1PIsct} can also be resummed in this way, with the result
\be
\lim_{\vec q\to0}\nabla_{q^i}\left(\frac{\langle\zeta_{\vec q}\zeta_{\vec k_1}\ldots\zeta_{\vec k_N}\rangle'}{P(q)}\right) = \frac{1}{2}\sum_{a=1}^N\left(-6\nabla_{k_a^i}+{\cal S}_a^i\right)\langle\zeta_{\vec k_1}\ldots\zeta_{\vec k_N}\rangle'~.
\ee
This is the~{\it conformal consistency relation}~\cite{Creminelli:2012ed}, now expressed in terms of the ${\cal S}_a$ operator. 

The procedure of translating the vertex Ward identities into a relation involving correlation functions will of great interest to us once we obtain the vertex Ward identities associated to higher soft limits.

\subsection{Double dilation}
\label{2softdils}

Having briefly reviewed what happens when a single momentum is taken to vanish, 
we now move on to the case of multiple soft external lines. At the level of $\Gamma$, things are relatively straightforward: the multi-soft Ward identities simply follow from repeated application of the identities~\eqref{1PIdilation} and~\eqref{1PIsct}.
The simplest example arises from taking another soft leg in~\eqref{1PIdilation}. The starting point is the identity~\eqref{1PIdilation} with $N\rightarrow N+1$:
\be
\lim_{\vec q_1\to0}\Gamma^{(N+2)}(\vec q_1,\vec q_2, \vec k_1,\cdots, \vec k_N) = \left(3-{\cal D}_N\right)\Gamma^{(N+1)}(\vec q_2,\vec k_1,\cdots, \vec k_N)~,
\label{order0double1}
\ee
where we have used the fact that ${\cal D}_{N+1} = {\cal D}_N$ when acting on $\Gamma^{(N+1)}$. Now we send $q_2\to 0$ and apply the dilation vertex identity again to obtain the {\it double dilation 1PI Ward identity}
\be
\label{order0double}
\lim_{\vec q_1, \vec q_2 \to 0}\Gamma^{(N+2)}(\vec q_1, \vec q_2, \vec k_1,\cdots, \vec k_N) = (3-{\cal D}_{N})(3-{\cal D}_{N})\Gamma^{(N)}(\vec k_1,\cdots, \vec k_N)~.
\ee
This represents a novel relation between the $(N+2)$-point 1PI vertex and the $N$-point vertex. 

This statement can equivalently be expressed in terms of correlation functions. For instance, consider the $N=2$ identity, relating the $4$-point function and the power spectrum.
The former can be written in terms of $\Gamma$ as
\begin{align}
\nonumber
\langle\zeta_{\vec q_1}\zeta_{\vec q_2}\zeta_{\vec k_1}\zeta_{\vec k_2}\rangle' = P(q_1)P(q_2)&P(k_1)P(k_2)\Big\{\Gamma^{(4)}(\vec q_1, \vec q_2, \vec k_1, \vec k_2)\\\nonumber
&~~~~~~+\Gamma^{(3)}(\vec q_1, \vec q_2, -\vec q)P(q)\Gamma^{(3)}(\vec k_1, \vec k_2, \vec q)\\\nonumber
&~~~~~~+\Gamma^{(3)}(\vec q_1, \vec k_1, -\vec q_1-\vec k_1)P(|\vec q_1+\vec k_1|)\Gamma^{(3)}(\vec q_2, \vec k_2, \vec q_1+\vec k_1)\\
&~~~~~~+\Gamma^{(3)}(\vec q_1, \vec k_2, -\vec q_1-\vec k_2)P(|\vec q_1+\vec k_2|)\Gamma^{(3)}(\vec k_1, \vec q_2, \vec q_1+\vec k_2)\Big\}~,
\label{4ptcorr}
\end{align}
where $\vec q \equiv \vec q_1+\vec q_2$. We wish to send both $\vec{q}_1,\vec{q}_2\rightarrow 0$. Although all three momenta in the vertex $\Gamma^{(3)}(\vec q_1, \vec q_2, -\vec q)$
go to zero in this limit, no hierarchy is assumed among the different modes, hence this is not a squeezed limit. For the vertex $\Gamma^{(3)}(\vec k_1, \vec k_2, \vec q)$, on the other hand,
the limit does correspond to a squeezed configuration $\vec q \ll \vec{k}_1,\vec{k}_2$. Using~\eqref{order0double1} and~\eqref{order0double}, we obtain
\begin{equation}
\lim_{\vec q_1, \vec q_2\to0}\frac{\langle\zeta_{\vec q_1}\zeta_{\vec q_2}\zeta_{\vec k_1}\zeta_{\vec k_2}\rangle'}{P(q_1)P(q_2)} = -\frac{\langle\zeta_{\vec q_1}\zeta_{\vec q_2}\zeta_{-\vec q}\rangle'}{P(q_1)P(q_2)}
\left(3+{\cal D}_1\right)P(k)+\Big(9+6{\cal D}_1+{\cal D}_1^2\Big)P(k)~.
\end{equation}
In terms of the dilation operator $\delta_{\cal D}  \equiv -3- {\cal D}_1$, this becomes
\be
\lim_{\vec q_1, \vec q_2\to0}\frac{\langle\zeta_{\vec q_1}\zeta_{\vec q_2}\zeta_{\vec k_1}\zeta_{\vec k_2}\rangle' }{P(q_1)P(q_2)}= \frac{\langle\zeta_{\vec q_1}\zeta_{\vec q_2}\zeta_{-\vec q}\rangle'}{P(q_1)P(q_2)}\delta_{\cal D}P(k)+\delta_{\cal D}^2P(k)~.
\label{2dilations}
\ee
This result generalizes to arbitrary $N$ in an obvious way:
\be
\lim_{\vec q_1, \vec q_2\to0}\frac{\langle\zeta_{\vec q_1}\zeta_{\vec q_2}\zeta_{\vec k_1}\cdots\zeta_{\vec k_N}\rangle' }{P(q_1)P(q_2)}= \frac{\langle\zeta_{\vec q_1}\zeta_{\vec q_2}\zeta_{-\vec q}\rangle'}{P(q_1)P(q_2)}\delta_{\cal D}\langle\zeta_{\vec k_1}\cdots\zeta_{\vec k_N}\rangle'+\delta_{\cal D}^2\langle\zeta_{\vec k_1}\cdots\zeta_{\vec k_N}\rangle'~,
\label{2dilationallN}
\ee 
with $\delta_{\cal D}  \equiv -3(N-1) - {\cal D}_N$. This precisely matches~\eqref{bwave2dil}, which was derived from background-wave arguments, thus confirming 
the equivalence between the two approaches.

\subsection{Dilation \& SCT}

Next we consider the ${\cal O}(q)$ contribution to the double-squeezed limit, which is related to performing a dilation and a SCT on the remaining hard legs. Na\"ively, there are two possible way to do this: we can either perform the dilation or the SCT first. However, as we will see these two choices lead to equivalent consistency relations, differing only by a commutator of the transformations. Additionally, they lead to the same result as the background wave approach.

To see this explicitly, we focus without loss of generality on the identity at ${\cal O}(q_1)$ and zeroth-order in $q_2$. The ambiguity in taking the double-soft limit now corresponds to whether $q_1$ or $q_2$ is taken to zero first.

\begin{itemize}

\item {\bf Send $\vec q_1 \rightarrow 0$, then  $\vec q_2 \rightarrow 0$.}\\
With this ordering, the relation is
\be
\lim_{\vec q_1, \vec q_2\to0}\nabla_{q_1}^i\Gamma^{(N+2)}\left(\vec q_1, \vec q_2, \vec k_1, \cdots, \vec k_{N-1}, -\vec q-\vec K\right) = \frac{1}{2}{\cal S}_N^i(3-{\cal D}_N)\Gamma^{(N)}\left(\vec k_1, \cdots, \vec k_{N-1}, -\vec K\right)~,
\label{interm1}
\ee
where $\vec q \equiv \vec q_1+\vec q_2$. Note that we have used the on-shell condition to express $\vec{k}_N$ in terms of the other $N-1$ momenta through $\vec K = \sum_{a=1}^{N-1}\vec k_a$. 
Since the right-hand side is on-shell, it only depends on $N-1$ variables. We can use this fact to eliminate the last term in the sum of ${\cal D}_N$ and ${\cal S}_N$, with the result\footnote{We henceforth suppress the arguments of $\Gamma$ for clarity, unless necessary.}
\be
\lim_{\vec q_1, \vec q_2\to0}\nabla_{q_1}^i\Gamma^{(N+2)}= \frac{1}{2}\left(3{\cal S}_{N-1}^i-{\cal S}_{N-1}^i{\cal D}_{N-1}\right)\Gamma^{(N)}~,
\label{orderqform1}
\ee

\item {\bf Send $\vec q_2 \rightarrow 0$, then  $\vec q_1 \rightarrow 0$.}\\
The identity in this case is
\be
\lim_{\vec q_1, \vec q_2\to0}\nabla_{q_1}^i\Gamma^{(N+2)} = \nabla_{q_1}^i(3-{\cal D}_N)\Gamma^{(N+1)} = (2-{\cal D}_N)\nabla_{q_1}^i\Gamma^{(N+1)}~,
\ee
where the last step follows from the $q_1$-derivative hitting $\vec q_1\cdot\vec\nabla_{q_1}$ part of ${\cal D}$. Using~the Ward identity~\eqref{1PIsct}, we obtain
\be
\lim_{\vec q_1, \vec q_2\to0}\nabla_{q_1}^i\Gamma^{(N+2)} = \frac{1}{2}\left(2{\cal S}^i_{N-1}-{\cal D}_{N-1}{\cal S}_{N-1}^i\right)\Gamma^{(N)}~.
\label{orderqform2}
\ee
Using the commutation relation $\left[{\cal S}_N^i, {\cal D}_N\right] = {\cal S}_N^i$ from the conformal algebra, we see that this agrees with~\eqref{orderqform1}.
Therefore, we see that in this formalism there is no preferred ordering, and we can use the identity in either form.

\end{itemize}

The vertex identity can be expressed in terms of correlation functions. Focusing once again on the $N=2$ identity relating the 4-point function to the power spectrum.
The starting point is~\eqref{4ptcorr} relating 4-point correlation function and vertex. Differentiating with respect to $q_1$, and using the identities~\eqref{1PIdilation},~\eqref{1PIsct},~\eqref{order0double}
and~\eqref{interm1}, we obtain the expression
\bea
\nonumber
\label{orderqinflationward}
\lim_{\vec q_1, \vec q_2\to0}\nabla^i_{q_1}\left(\frac{\langle\zeta_{\vec q_1}\zeta_{\vec q_2}\zeta_{\vec k_1}\zeta_{\vec k_2}\rangle'}{P(q_1)P(q_2)} \right)&=& \frac{1}{2}\delta_{{\cal K}^i}\delta_{\cal D}P(k)+\frac{1}{2}\left(\frac{\langle\zeta_{\vec q_1}\zeta_{\vec q_2}\zeta_{-\vec q}\rangle'}{P(q_1)P(q_2)}\right)\delta_{{\cal K}^i}P(k)\\
&~& +~\nabla_{q_1}^i\left(\frac{\langle\zeta_{\vec q_1}\zeta_{\vec q_2}\zeta_{-\vec q}\rangle'}{P(q_1)P(q_2)}\right)\delta_{\cal D}P(k)~,
\eea
where $\delta_{{\cal K}^i} = -6\nabla_k^i+{\cal S}_1^i$, which is just the normal SCT operator acting on the 2-point function. 
As before, this identity has an obvious generalization to $N$-point functions:
\bea
\label{dil_sct}
\nonumber
\lim_{\vec q_1, \vec q_2\to0}\nabla^i_{q_1}\left(\frac{\langle\zeta_{\vec q_1}\zeta_{\vec q_2}\zeta_{\vec k_1}\cdots\zeta_{\vec k_N}\rangle'}{P(q_1)P(q_2)} \right)&=& \frac{1}{2}\delta_{{\cal K}^i}\delta_{\cal D}\langle\zeta_{\vec k_1}\cdots\zeta_{\vec k_N}\rangle'+\frac{1}{2}\left(\frac{\langle\zeta_{\vec q_1}\zeta_{\vec q_2}\zeta_{-\vec q}\rangle'}{P(q_1)P(q_2)}\right)\delta_{{\cal K}^i}\langle\zeta_{\vec k_1}\cdots\zeta_{\vec k_N}\rangle'\\
&~&+~ \nabla_{q_1}^i\left(\frac{\langle\zeta_{\vec q_1}\zeta_{\vec q_2}\zeta_{-\vec q}\rangle'}{P(q_1)P(q_2)}\right)\delta_{\cal D}\langle\zeta_{\vec k_1}\cdots\zeta_{\vec k_N}\rangle'~,
\eea
which precisely matches the $q_1$-derivative of the identity~\eqref{dilthensctconsis}. The equivalence of the differential form of identities of this type and the ``re-summed" form was shown in~\cite{Creminelli:2012qr,Goldberger:2013rsa}, and those arguments can be adapted straightforwardly to the present case.

\subsection{Double SCT}

Finally we consider the ${\cal O}(q^2)$ contribution to the double-squeezed limit, corresponding to two SCTs. Since two SCTS commute, we choose
to send $\vec{q}_2\rightarrow 0$ first without loss of generality. Applying~\eqref{1PIsct}, we obtain
\bea
\nonumber
& & \lim_{\vec q_1, \vec q_2\to0}\nabla_{q_1}^i\nabla_{q_2}^j\Gamma^{(N+2)}\left(\vec q_1,\vec q_2,\vec k_1, \cdots, \vec k_{N-1}, -\vec q_1-\vec q_2-\vec K\right) \\
& & ~~~~~~~~~~~~~~~~~~~~~~~~~ = \lim_{\vec q_1\to0}\frac{1}{2}\nabla_{q_1}^i{\cal S}_N^j\Gamma^{(N+1)}\left(\vec q_1,\vec k_1, \cdots, \vec{k}_{N-1},-\vec q_1-\vec K\right)~.
\eea
The right-hand side generates two terms: one where $\nabla_q^i$ hits the SCT generator, and one where it hits $\Gamma^{(N+1)}$. Explicitly, the result is
\bea
\nonumber
\lim_{\vec q_1, \vec q_2\to0}\nabla_{q_1}^i\nabla_{q_2}^j\Gamma^{(N+2)} &= &\lim_{q_1\to0}\frac{1}{2}\left(\delta^{ij}\nabla^2_{q_1}-2\nabla^i_{q_1}\nabla^j_{q_1}\right)\Gamma^{(N+1)}\left(\vec q_1,\vec k_1, \cdots, -\vec q_1-\vec K\right)\\
\nonumber
& & +\lim_{\vec q_1\to0}\frac{1}{2}{\cal S}_N^j\nabla^i_{q_1}\Gamma^{(N+1)}\left(\vec q_1,\vec k_1, \cdots, -\vec q_1-\vec K\right) \\
\nonumber
&= & \lim_{\vec q_1\to0}\frac{1}{2}\left(\delta^{ij}\nabla^2_{q_1}-2\nabla^i_{q_1}\nabla^j_{q_1}\right)\Gamma^{(N+1)}\left(\vec q_1,\vec k_1, \cdots, -\vec q_1-\vec K\right)\\
&& + \frac{1}{4}{\cal S}_{N-1}^j{\cal S}^i_{N-1}\Gamma^{(N)}\left(\vec k_1, \cdots, \vec k_N\right)~,
\label{orderq2vertex}
\eea
where in the last step we have again used the single-soft SCT identity~\eqref{1PIsct}.\footnote{The use the SCT Ward identity is clearly justified for terms in ${\cal S}$ involving derivatives with respect to $k_a\neq q$. Less obvious
is the term in ${\cal S}$ involving derivatives with respect to $q_1$. However, this term is multiplied by $q_1$ and hence gives a vanishing contribution in the soft limit.} 

This identity contains two terms: the ${\cal S}_{N-1}^j{\cal S}^i_{N-1}\Gamma^{(N)}$ term represents two SCTs acting on the $N$-point vertex, which was expected; the other term, involving $\partial^2_{q_1}\Gamma^{(N+1)}$,
cannot be further reduced to an $N$-point vertex. This is not surprising, as mentioned earlier, because ${\cal O}(q^2)$ terms are of the same order as the curvature of spatial slices, which is a physical effect and cannot be removed by a diffeomorphism. This curvature contribution cannot be fixed in terms of a symmetry transformation on the lower-point vertex. What~\eqref{orderq2vertex} provides is a highly non-trivial relation between the ${\cal O}(q^2)$ part of the $(N+2)$-point function in the double-soft limit, the ${\cal O}(q^2)$ part of the soft limit of the $(N+1)$-point  and a symmetry transformation of the $N$-point function with all hard momenta. 

Again, we want to translate this 1PI identity into a statement about more familiar correlation functions of $\zeta$. This is a rather intricate task, but by taking derivatives of~\eqref{4ptcorr} with respect to $\vec q_1$ and $\vec q_2$, employing the identities ~\eqref{1PIdilation},~\eqref{1PIsct},~\eqref{order0double},~\eqref{interm1} and ~\eqref{orderq2vertex}, and doing a sizable amount of algebra, one can obtain the relation
\begin{align}
\nonumber
\label{orderq2wardidentity}
\lim_{\vec q_1, \vec q_2\to 0}\nabla_{q_1}^j\nabla_{q_2}^i\left(
\frac{\langle\zeta_{\vec q_1}\zeta_{\vec q_2}\zeta_{\vec k_1}\zeta_{\vec k_2}\rangle'}{P(q_1)P(q_2)}\right)  =
&~\nabla_{q_1}^j\nabla_{q_2}^i\left(\frac{\langle\zeta_{\vec q_1}\zeta_{\vec q_2}\zeta_{-\vec q}\rangle'}{P(q_1)P(q_2)}\right)\delta_{\cal D}P(k)\\\nonumber
&+
\frac{1}{2}\nabla_{q_1}^j\left(\frac{\langle\zeta_{\vec q_1}\zeta_{\vec q_2}\zeta_{-\vec q}\rangle'}{P(q_1)P(q_2)}\right)\delta_{{\cal K}^i}P(k)\\\nonumber
&+\frac{1}{2}\nabla_{q_2}^i\left(\frac{\langle\zeta_{\vec q_1}\zeta_{\vec q_2}\zeta_{-\vec q}\rangle'}{P(q_1)P(q_2)}\right)\delta_{{\cal K}^j}P(k)\\\nonumber
&+ \frac{1}{4}\delta_{{\cal K}^i} \delta_{{\cal K}^j}P(k) \\\nonumber
 &+\lim_{\vec p\to 0}\frac{1}{2}(\delta^{ij}\nabla_p^2-2\nabla_p^i\nabla_p^j)\frac{\langle\zeta_{\vec p}\zeta_{\vec k_1}\zeta_{\vec k_2}\rangle'}{P(p)}\\
 &+\left(\frac{\langle\zeta_{\vec q_1}\zeta_{\vec q_2}\zeta_{-\vec q}\rangle'}{P(q_1)P(q_2)}\right)\lim_{\vec p\to 0} \nabla_p^i\nabla_p^j\frac{\langle\zeta_{\vec p}\zeta_{\vec k_1}\zeta_{\vec k_2}\rangle'}{P(p)}~.
\end{align}
Which relates the $\vec q_1\cdot \vec q_2$ part of the squeezed limit of the $4$-point function to a squeezed limit of the $3$-point function and the power spectrum. This identity generalizes in a straightforward way to $(N+2)$-point functions:
\begin{align}
\nonumber
\label{orderq2wardidentity}
\lim_{\vec q_1,\vec q_2\to 0}\nabla_{q_1}^j\nabla_{q_2}^i\left(
\frac{\langle\zeta_{\vec q_1}\zeta_{\vec q_2}\zeta_{\vec k_1}\cdots\zeta_{\vec k_N}\rangle'}{P(q_1)P(q_2)}\right)  =
&~\nabla_{q_1}^j\nabla_{q_2}^i\left(\frac{\langle\zeta_{\vec q_1}\zeta_{\vec q_2}\zeta_{-\vec q}\rangle'}{P(q_1)P(q_2)}\right)\delta_{\cal D}\langle\zeta_{\vec k_1}\cdots\zeta_{\vec k_N}\rangle'\\\nonumber
&+
\frac{1}{2}\nabla_{q_1}^j\left(\frac{\langle\zeta_{\vec q_1}\zeta_{\vec q_2}\zeta_{-\vec q}\rangle'}{P(q_1)P(q_2)}\right)\delta_{{\cal K}^i}\langle\zeta_{\vec k_1}\cdots\zeta_{\vec k_N}\rangle'\\\nonumber
&+\frac{1}{2}\nabla_{q_2}^i\left(\frac{\langle\zeta_{\vec q_1}\zeta_{\vec q_2}\zeta_{-\vec q}\rangle'}{P(q_1)P(q_2)}\right)\delta_{{\cal K}^j}\langle\zeta_{\vec k_1}\cdots\zeta_{\vec k_N}\rangle'\\\nonumber
&+ \frac{1}{4}\delta_{{\cal K}^i} \delta_{{\cal K}^j}\langle\zeta_{\vec k_1}\cdots\zeta_{\vec k_N}\rangle' \\\nonumber
 &+\lim_{\vec p\to 0}\frac{1}{2}(\delta^{ij}\nabla_p^2-2\nabla_p^i\nabla_p^j)\frac{\langle\zeta_{\vec p}\zeta_{\vec k_1}\cdots\zeta_{\vec k_N}\rangle'}{P(p)}\\
 &+\left(\frac{\langle\zeta_{\vec q_1}\zeta_{\vec q_2}\zeta_{-\vec q}\rangle'}{P(q_1)P(q_2)}\right)\lim_{p\to 0} \nabla_p^i\nabla_p^j\frac{\langle\zeta_{\vec p}\zeta_{\vec k_1}\cdots\zeta_{\vec k_N}\rangle'}{P(p)}~.
\end{align}
Notice that -- as was the case for the ${\cal O}(q)$ identity -- the expression is the derivative with respect to $\vec q_1$ and $\vec q_2$ of the expression~\eqref{2scts}, confirming the equivalence between the methods of deriving these identities.

\section{Checks of double-soft identities}
\label{corrcheck}

In this Section we perform some checks of our consistency relations using explicit inflationary correlation functions. Since our identities constrain correlation functions in the double-soft limit, we must consider at least a $4$-point function on the left-hand side. Unfortunately, there are few 4-point computations in the literature. Our checks are therefore limited to two cases where such computations are tractable: models with resonant non-Gaussianities and models with small speed of sound. 

\subsection{Resonant non-Gaussianity}
\label{1piresonantcheck}

Resonant non-Gaussianity~\cite{Flauger:2009ab,Flauger:2010ja} arises in models with a periodic modulation of the inflationary potential~\cite{Chen:2008wn,McAllister:2008hb,Kaloper:2008fb}. They offer a nice arena to check our consistency relations because it is possible to compute $N$-point functions, in principle for arbitrary $N$~\cite{Leblond:2010yq,Behbahani:2011it}. In this case, it is simplest to check the Ward identities at the level of the 1PI action.
Considering only contact contributions,\footnote{It is consistent to consider contact diagrams only, since these give the only contributions proportional to the first power of the frequency of modulation of the potential. Therefore, they must satisfy the consistency relations by themselves~\cite{Creminelli:2012ed, Leblond:2010yq}.} the $N$-point correlation function is given by~\cite{Creminelli:2012ed, Leblond:2010yq, Behbahani:2011it}
\be
\langle \zeta_{\vec k_1}\zeta_{\vec k_2}\cdots\zeta_{\vec k_N}\rangle_{\rm rnG}' =-2 \left(\frac{1}{2\epsilon}\right)^{N/2}\frac{H^{2N-4}}{\prod_{i=1}^N2k_i^3}~{\rm Im}\int_{-\infty}^0\frac{\rd\eta}{\eta^4}V^{(N)}(\phi)\prod_{a=1}^N(1-ik_a\eta)e^{ik_a\eta}~,
\ee
where $V^{(N)}(\phi) \equiv \frac{{\rm d}^NV}{{\rm d}\phi^N}$. The power spectrum is given by
\be
P(k) = \frac{H^2}{4\epsilon k^3}~.
\ee
Since this $N$-point function only contains contact interactions, we have the relation
\bea
\nonumber
\Gamma_{\rm rnG}^{(N)}(\vec k_1, \vec k_2, \ldots, \vec k_N) &=& \frac{\langle \zeta_{k_1}\zeta_{k_2}\cdots\zeta_{k_N}\rangle_{\rm rnG}'}{P(k_1)\cdots P(k_N)} \\
&=& -\frac{2}{H^4} \left(2\epsilon\right)^{N/2}~{\rm Im}\int_{-\infty}^0\frac{\rd\eta}{\eta^4}V^{(N)}(\phi)\prod_{a=1}^N(1-ik_a\eta)e^{ik_a\eta}~.
\label{resonant1pi}
\eea

\begin{itemize} 

\item {\bf Dilation}: Using the identity
\be
V^{(N)}(\phi) = \frac{\rd}{\rd\phi}V^{(N-1)}(\phi) = \left(\frac{1}{2\epsilon}\right)\eta\frac{\rd}{\rd\eta}V^{(N-1)}(\phi)~,
\label{resonantNGibp}
\ee
we can integrate by parts to obtain\footnote{The integration by parts also generates a boundary term. However, once we regulate the integral, we see that the boundary term is real and therefore can be discarded.}
\be
\lim_{\vec{k}_1 \to 0}\Gamma_{\rm rnG}^{(N)}= \frac{2}{H^4} \left(2\epsilon \right)^{(N-1)/2}~ {\rm Im}\int_{-\infty}^0\rd\eta V^{(N-1)}(\phi)\frac{\rd}{\rd\eta}\left(\frac{1}{\eta^3}\prod_{a=2}^N(1-ik_a\eta)e^{ik_a\eta}\right)~.
\label{resonantngdil1}
\ee
The time derivative gives 
\bea
\nonumber
\frac{\rd}{\rd\eta}\left(\frac{1}{\eta^3}\prod_{a=2}^N(1-ik_a\eta)e^{ik_a\eta}\right) &=& \frac{1}{\eta^4} \left(-3 +\sum_{i=2}^{N}\vec k_a\cdot\vec\nabla_{k_a}\right)\prod_{i=2}^N(1-ik_i\eta)e^{ik_i\eta}\\
&=& -\frac{1}{\eta^4} \left(3 - {\cal D}_{N-1}\right)\prod_{i=2}^N(1-ik_i\eta)e^{ik_i\eta}\,.
\label{timederdil}
\eea
The terms inside the integral combine to give $\Gamma^{(N-1)}$, and it immediately follows that
\be
\lim_{\vec{k}_1 \to 0}\Gamma^{(N)}_{\rm rnG}(\vec k_1, \vec k_2, \cdots,\vec  k_N) = (3-{\cal D}_{N-1})\Gamma_{\rm rnG}^{(N-1)}(\vec k_2, \cdots, \vec k_N)~.
\label{RNGdil}
\ee
This establishes the 1PI dilation identity~\eqref{1PIdilation} in theories with resonant non-Gaussianities.

\item {\bf SCT}: Next we check the single-soft SCT relation~\eqref{1PIsct}. Taking the derivative of~\eqref{resonant1pi}, we obtain 
\be
\lim_{\vec k_1\to0}\nabla_{k_1}^i\Gamma^{(N)}_{\rm rnG} =  \frac{2}{H^4} \hat k_N^i\left(2\epsilon \right)^{N/2}~{\rm Im}\int\frac{\rd\eta}{\eta^2}V^{(N)}(\phi)e^{i\hat k_N\eta}\prod_{a=2}^{N-1}(1-ik_a\eta)e^{ik_a\eta}~,
\ee
where we have defined $\hat k^i_N = -k_2^i-\ldots-k^i_{N-1}$. (This is just the on-shell condition with $\vec k_1\to 0$.)
Once again we use~\eqref{resonantNGibp} and integrate by parts to rewrite this expression as
\be
\lim_{\vec k_1\to0}\nabla_{k_1}^i\Gamma^{(N)}_{\rm rnG} = -\frac{2}{H^4} \hat k_N^i \left(2\epsilon \right)^{(N-1)/2}{\rm Im}\int\rd\eta V^{(N-1)}(\phi) \frac{\rd}{\rd\eta}\left(\frac{e^{i\hat k_N\eta}}{\eta}\prod_{a=2}^{N-1}(1-ik_a\eta)e^{ik_a\eta}\right)~,
\label{LHSorderq}
\ee
After a considerable amount of algebra, similar to~\eqref{timederdil}, the time-derivative inside the integral can be repackaged as
\be
\hat k_N^i \frac{\rd}{\rd\eta}\left(\frac{e^{i\hat k_N\eta}}{\eta}\prod_{i=2}^{N-1}(1-ik_i\eta)e^{ik_i\eta}\right) = \frac{1}{2\eta^4}{\cal S}_{N-2}^i\prod_{a=2}^N(1-i\hat k_a\eta)^{ik_a\eta}\,,
\ee
where ${\cal S}_{N-2} = \sum_{a=2}^{N-1}{\cal S}_a^i$. As before, the integral becomes $\Gamma^{(N-1)}$, with the result
\be
\lim_{\vec k_1\to0}\nabla_{k_1}^i\Gamma^{(N)}_{\rm rnG}(\vec k_1, \vec k_2, \cdots, \vec k_N) = \frac{1}{2}{\cal S}_{N-1}^i\Gamma^{(N-1)}_{\rm rnG}(\vec k_2, \cdots, \vec k_N)~.
\label{RNGSCT}
\ee
This establishes the 1PI SCT identity~\eqref{1PIdilation} in theories with resonant non-Gaussianities.

\end{itemize}

\noindent Since the 1PI identities~\eqref{RNGdil} and~\eqref{RNGSCT} have been verified for arbitrary $N$, by iteration the multi-soft relations~\eqref{order0double},~\eqref{orderqform1} and~\eqref{orderq2vertex} hold as well. Additionally, this means that the identities in terms of correlation functions hold. For example, using the first line of~\eqref{resonant1pi}, it is straightforward to show that
\be
\lim_{\vec q_1, \vec q_2 \to 0}\frac{\langle\zeta_{\vec q_1}\zeta_{\vec q_2}\zeta_{\vec k_1}\cdots\zeta_{\vec k_N}\rangle'_{\rm rnG}}{P(q_1)P (q_2)} =   \delta_{\cal D}^2 \langle\zeta_{\vec k_1}\cdots\zeta_{\vec k_N}\rangle'_{\rm rnG}~.
\ee
Notice that in this case, only the term with two dilations contributes at leading order, this is because in these models, all exchange contributions are sub-leading in powers of the modulation of the potential.

\subsection{Small speed of sound}
\label{smallcs}

Another arena where we can check our double-soft identities is in models with $c_s\ll 1$. To do this, we use the full 4-point correlation function, calculated in~\cite{Chen:2009bc,Arroja:2009pd}
and summarized in the Appendix. Since the result is rather complicated, below we will just quote the double-squeezed results. 

Note that such a check is rather non-trivial, and crucially involves the presence of the terms involving gradients of the $(N+1)$-point function in~\eqref{orderq2wardidentity}. This can be seen from the scaling of the various correlation functions with respect to $c_s$: the 4-point function scales like $c_s^{-7}$ as $c_s \to 0$, while the 3pt. function scales as $c_s^{-4}$ and the power spectrum scales as $c_s^{-1}$. The na\"ive double-soft limit would relate the 4pt. function to two terms: one involving 3 powers of the power spectrum, and one involving the 3pt. function times one power spectrum. In both cases, it is clear that the powers of $c_s$ cannot match. The resolution is of course the presence of the additional terms proportional to squeezed 3pt. functions, for which the powers of $c_s$ {\it do} match the 4pt. function.

In these models, the power spectrum is exactly scale-invariant,
\be
P(k) = \( \frac{H^2}{4M_{\rm Pl}^2c_s\epsilon} \)\frac{1}{k^3}~,
\ee
and hence is annihilated by both $\delta_{\cal D}$ and $\delta_{{\cal K}^i}$. It follows that the right-hand side of the double dilation~\eqref{2dilations} and dilation-SCT~\eqref{orderqinflationward} identities vanish identically. The left-hand sides of these relations also vanish, since as we will see the 4-point function starts at ${\cal O}(q^2)$. The only non-trivial check is the double-SCT relation~\eqref{orderq2wardidentity}, which for scale-invariant power spectrum
reduces to
\bea
\lim_{\vec q_1,\vec q_2\to 0}\nabla_{q_1}^j\nabla_{q_2}^i\left(
\frac{\langle\zeta_{\vec q_1}\zeta_{\vec q_2}\zeta_{\vec k_1}\zeta_{\vec k_2}\rangle'}{P(q_1)P(q_2)}\right)  =
 \lim_{\vec q\to 0} \Bigg\{\frac{1}{2}\delta^{ij}\nabla_q^2 + \left(\frac{\langle\zeta_{\vec q_1}\zeta_{\vec q_2}\zeta_{-\vec q}\rangle'}{P(q_1)P(q_2)} -1\right) \nabla_q^i\nabla_q^j\Bigg\}\frac{\langle\zeta_{\vec q}\zeta_{\vec k}\zeta_{-\vec k}\rangle'}{P(q)}\,.
\label{reducedoq2ward}
\eea

There are two types of contributions to the 4-point function, those coming from contact interactions and those coming from scalar exchange:
\be
\langle \zeta_{\vec q_1} \zeta_{\vec q_2} \zeta_{\vec k_1} \zeta_{\vec k_2} \rangle'= \langle \zeta_{\vec q_1} \zeta_{\vec q_2} \zeta_{\vec k_1} \zeta_{\vec k_2} \rangle'_{\rm cont.}+\langle \zeta_{\vec q_1} \zeta_{\vec q_2} \zeta_{\vec k_1} \zeta_{\vec k_2} \rangle'_{\rm exc.}
\ee
In the double-squeezed limit, the contact contribution gives
\be
\lim_{\vec q_1, \vec q_2\to0}\langle \zeta_{\vec q_1} \zeta_{\vec q_2} \zeta_{\vec k_1} \zeta_{\vec k_2} \rangle'_{\rm cont.}= P(q_1)P(q_2) P(k) \(1 -\frac{1}{c_s^2} \) \( \frac 3 4 \frac{\vec q_1 \cdot \vec q_2}{k^2} + \frac 52 \frac{(\vec k \cdot \vec q_1)(\vec k \cdot \vec q_2)}{k^4}\) \;.
\label{cont}
\ee
The exchange diagram, meanwhile, is expected to factorize:
\be
\lim_{\vec q_1, \vec q_2\to0}\langle \zeta_{\vec q_1} \zeta_{\vec q_2} \zeta_{\vec k_1} \zeta_{\vec k_2} \rangle'_{\rm exc.}
= \lim_{\vec q\to0}\langle \zeta_{\vec q_1} \zeta_{\vec q_2} \zeta_{-\vec q} \rangle' \frac{\langle \zeta_{\vec q} \zeta_{\vec k_1} \zeta_{\vec k_2} \rangle'}{P(q)} 
\label{exch1}
\ee
Assuming scale invariance, the last factor vanishes up to ${\cal O}(q^2)$ by the usual consistency relations. At order $q^2$, however,
we have the  ``curvature" consistency relation~\cite{Creminelli:2013cga}
\be
\lim_{\vec q\to0}~\langle \zeta_{\vec q} \zeta_{\vec k_1} \zeta_{\vec k_2} \rangle' = P(q)P(k) \(1 -\frac{1}{c_s^2}  \) \( 2\frac{q^2}{k^2} - \frac 54 \frac{(\vec k \cdot \vec q)^2}{k^4}\) \;.
\label{squeezed3pt}
\ee
Substituting into~\eqref{exch1}, the exchange contribution becomes
\be
\lim_{\vec q_1, \vec q_2\to0}\langle \zeta_{\vec q_1} \zeta_{\vec q_2} \zeta_{\vec k_1} \zeta_{\vec k_2} \rangle'_{\rm exc.} = P(k)\(1 -\frac{1}{c_s^2}  \) \( 2\frac{q^2}{k^2} - \frac 54 \frac{(\vec k \cdot \vec q)^2}{k^4}\) \langle \zeta_{\vec q_1} \zeta_{\vec q_2} \zeta_{-\vec q} \rangle' \;.
\label{exch}
\ee
This expectation is borne out by explicitly calculating the double-squeezed limit of the $4$-point function. Both contact~\eqref{cont} and exchange~\eqref{exch} contributions start at order $q^2$, as claimed. Taking derivatives with respect to $q_1$ and $q_2$, we obtain\footnote{The second term gets a factor of 2 from the fact that $q^2 \supset 2q_1\cdot q_2$.}
\bea
\nonumber
\lim_{\vec q_1,\vec q_2\to 0}\nabla_{q_1}^j\nabla_{q_2}^i\left(\frac{\langle\zeta_{\vec q_1}\zeta_{\vec q_2}\zeta_{\vec k_1}\zeta_{\vec k_2}\rangle'}{P(q_1)P(q_2)}\right) &=&P(k)\left(1-\frac{1}{c_s^2}\right)\Bigg\{\left(\frac{3}{4}\frac{\delta^{ij}}{k^2}+\frac{5}{2}\frac{k^ik^j}{k^4}\right)\\
& &~~~~~~ +  \left(4\frac{\delta^{ij}}{k^2}-\frac{5}{2}\frac{k^ik^j}{k^4}\right)\frac{\langle\zeta_{\vec q_1}\zeta_{\vec q_2}\zeta_{-\vec q}\rangle'}{P(q_1)P(q_2)}\Bigg\}\,.
\label{oq2wardLHS}
\eea

Meanwhile, for the right-hand side of~\eqref{reducedoq2ward}, the 3-point function is written explicitly in the Appendix~\eqref{3ptsmallcs}, its squeezed limit is~\eqref{squeezed3pt}. Using this, we obtain 
\be
\lim_{\vec q\to 0}\nabla_q^i\nabla_q^j\frac{\langle\zeta_{\vec q}\zeta_{\vec k}\zeta_{-\vec k}\rangle' }{P(q)}= P(k)\left(1-\frac{1}{c_s^2}\right)\left(4\frac{\delta^{ij}}{k^2}-\frac{5}{2}\frac{k^i k^j}{k^4}\right)~.
\ee
This leads to the following expression for the right-hand side of~\eqref{reducedoq2ward}:
\begin{align}
\nonumber
\lim_{\vec q\to 0} \Bigg\{\frac{1}{2}\delta^{ij}\nabla_q^2 + \left(\frac{\langle\zeta_{\vec q_1}\zeta_{\vec q_2}\zeta_{-\vec q}\rangle'}{P(q_1)P(q_2)} -1\right) \nabla_q^i\nabla_q^j\Bigg\}\frac{\langle\zeta_{\vec q}\zeta_{\vec k}\zeta_{-\vec k}\rangle'}{P(q)} &= P(k)\left(1-\frac{1}{c_s^2}\right)\Bigg\{\left(\frac{3}{4}\frac{\delta^{ij}}{k^2}+\frac{5}{2}\frac{k^ik^j}{k^4}\right)\\
&+  \left(4\frac{\delta^{ij}}{k^2}-\frac{5}{2}\frac{k^ik^j}{k^4}\right)\frac{\langle\zeta_{\vec q_1}\zeta_{\vec q_2}\zeta_{-\vec q}\rangle'}{P(q_1)P(q_2)}\Bigg\},
\label{oq2wardRHS}
\end{align}
which precisely agrees with~\eqref{oq2wardLHS}.

\section{$N> 2$ Soft external legs}
\label{Nsoft}

In this Section, we generalize the derivation to multiple external soft legs. This is completely straightforward at the level of the 1PI action -- all we need to do is apply the dilation~\eqref{1PIdilation} and SCT~\eqref{1PIsct} identities repeatedly. For instance, the multi-dilation identity is
\be
\lim_{\vec q_1,\cdots, \vec q_N \to 0}\Gamma^{(M+N)}(\vec q_1, \cdots, \vec q_N, \vec k_1,\cdots, \vec k_M) = (3-{\cal D}_{M})^N\Gamma^{(M)}(\vec k_1,\cdots, \vec k_M)~,
\label{multidil}
\ee
which just tells us that the dilation operator $(3-{\cal D}_M)$ acts $N$ times on the remaining hard modes in the vertex. Note that for scale-invariant theories, there is an analogue of Adler's zero~\cite{Adler:1964um}: by taking any number of soft legs, the right hand side will vanish, being proportional to a scale variation of some lower order correlation function.

\begin{figure}
\centering
\begin{subfigure}{}
\includegraphics[width=2.2in]{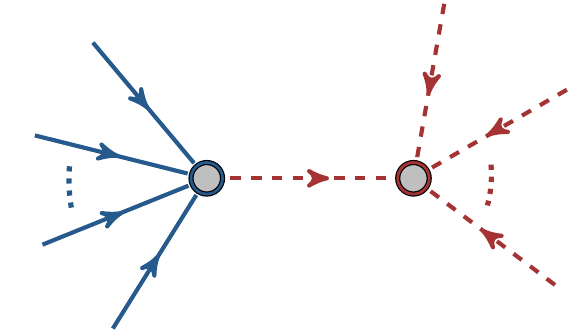}
\end{subfigure}
\begin{subfigure}{}
\includegraphics[width=2.2in]{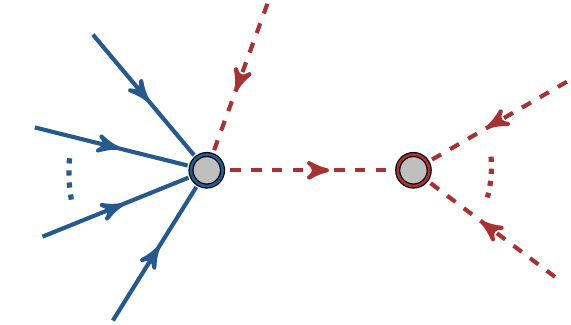}
\end{subfigure}
\begin{subfigure}{}
\includegraphics[width=1.7in]{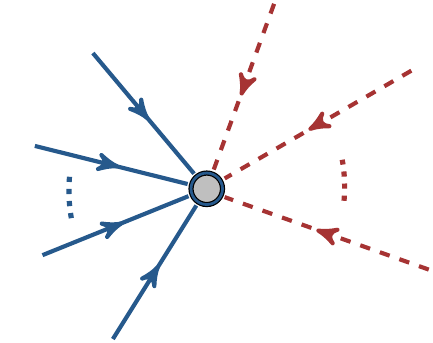}
\end{subfigure}
\caption{\label{Ndilationfig}{\small Contributions to the $N$-squeezed limit~\eqref{Ndilations}. {\it Left}: Contribution where the all soft modes to combine to factorize the diagram via a soft internal line. {\it Middle}: Contribution all but one of the soft modes combine to factorize the diagram, while the last mode comes from the same vertex as the hard mode, causing the remaining hard modes to feel two dilations. {\it Right}: Diagram where all soft modes come from the same vertex as the hard lines, causing the correlation function to feel $N$ dilations.}}
\end{figure}

Rephrasing such identities in terms of correlation functions is a slightly tricky task, hence we will only discuss the above multi-dilation identity. We can gain some insight into the problem
by thinking diagrammatically, in the same was as we did in Sec.~\ref{2softdils}. Various diagrams contribute to the $N$-soft limit -- see Fig.~\ref{Ndilationfig}. The simplest diagram (left panel)
is where all soft modes add up to a small momentum, thereby causing the diagram to factorize. This leads to a contribution to the soft limit of the form (in the following expressions, $\vec q_a \ll \vec k_a$, but all the $\vec q_a$ are of the same order) :
\be
\langle\zeta_{\vec q_1}\cdots\zeta_{\vec q_N}\zeta_{-\vec q}\rangle'\frac{\langle\zeta_{\vec q}\zeta_{\vec k_1}\cdots\zeta_{\vec k_M}\rangle' }{P(q)}= \langle\zeta_{\vec q_1}\cdots\zeta_{\vec q_N}\zeta_{-\vec q}\rangle' \delta_{\cal D}\langle\zeta_{\vec k_1}\cdots\zeta_{\vec k_M}\rangle'~.
\ee
The next simplest situation, shown on the middle panel, is where $(N-1)$ of the soft momenta add up to a small momentum and factorize the diagram, while the remaining soft leg emerges from the same vertex as the hard modes. This leads to a contribution of the form:
\be
\langle\zeta_{\vec q_1}\cdots\zeta_{\vec q_{N-1}}\zeta_{-\vec q}\rangle'\frac{\langle\zeta_{\vec q}\zeta_{\vec q_N}\zeta_{\vec k_1}\cdots\zeta_{\vec k_M}\rangle'}{P(q)} =  P (q_N)\langle\zeta_{\vec q_1}\cdots\zeta_{\vec q_{N-1}}\zeta_{-\vec q}\rangle' \delta_{\cal D}^2 \langle\zeta_{\vec k_1}\cdots\zeta_{\vec k_M}\rangle'~.
\ee
Similarly, we can have $L < N$ of the modes all adding up to a small momentum while the remaining modes come from the same vertex as the hard legs, leading to a contribution of the form
\be
\left(\prod_{i=N-L}^{N}P(q_i)\right) \langle\zeta_{\vec k_{N-L}}\cdots\zeta_{\vec k_L}\zeta_{-\vec q}\rangle'\delta_{\cal D}^{(N-L+1)} \langle\zeta_{\vec k_1}\cdots\zeta_{\vec k_M}\rangle'
\ee
Finally, all $N$ soft legs can emerge from the same vertex as the hard modes, as shown on the right panel. This amounts to $N$ dilations of the hard modes.
Putting everything together, we expect the $N$-dilation Ward identity to take the form 
\begin{align}
\nonumber
\lim_{\vec q_1,\cdots \vec q_N \to 0}\frac{\langle\zeta_{\vec q_1}\cdots\zeta_{\vec q_N}\zeta_{\vec k_1}\cdots\zeta_{\vec k_M}\rangle'}{P(q_1)\cdots P(q_N)}  =& ~\frac{\langle\zeta_{\vec q_1}\cdots\zeta_{\vec q_N}\zeta_{-\vec q}\rangle'}{P(q_1)\cdots P(q_N)} \delta_{\cal D}\langle\zeta_{\vec k_1}\cdots\zeta_{\vec k_M}\rangle'\\\nonumber
& +  \frac{\langle\zeta_{\vec q_1}\cdots \zeta_{\vec q_{N-1}}\zeta_{-\vec q}\rangle'}{P(q_1)\cdots P(q_{N-1})} \delta_{\cal D}^2\langle\zeta_{\vec k_1}\cdots\zeta_{\vec k_M}\rangle' + \cdots \\\nonumber
&+ \frac{\langle\zeta_{\vec q_{1}}\cdots\zeta_{\vec q_L}\zeta_{-\vec q}\rangle'}{P(q_1)\cdots P(q_{L})} \delta_{\cal D}^{(N-L+1)} \langle\zeta_{\vec k_1}\cdots\zeta_{\vec k_M}\rangle' + \cdots\\
&+ \delta_{\cal D}^N\langle\zeta_{\vec k_1}\cdots\zeta_{\vec k_M}\rangle'~,
\label{Ndilations}
\end{align}
which can be succinctly expressed as
\be
\lim_{\vec q_1,\cdots \vec q_N \to 0}\frac{\langle\zeta_{\vec q_1}\cdots\zeta_{\vec q_N}\zeta_{\vec k_1}\cdots\zeta_{\vec k_M}\rangle'}{P(q_1)\cdots P(q_N)} = \sum_{L =1}^{N}  \frac{\langle\zeta_{\vec q_{1}}\cdots\zeta_{\vec q_{L}}\zeta_{-\vec q}\rangle'}{P(q_1)\cdots P(q_{L})} \delta_{\cal D}^{(N-L+1)} \langle\zeta_{\vec k_1}\cdots\zeta_{\vec k_M}\rangle'~.
\ee
The generalization at higher powers in the soft momenta, corresponding to mixed dilation/SCT identities, is tedious but straightforward.





\section{Large scale structure consistency relations}
\label{LSS}

So far we have considered the case of single-field inflation and showed how the standard consistency relations for a single soft momentum generalize to the case of multiple soft momenta. In this Section we want to focus on the recently derived consistency relations for Large Scale Structure~\cite{Creminelli:2013mca, Creminelli:2013poa, Creminelli:2013nua,Valageas:2013cma,Peloso:2013zw,Kehagias:2013yd,Peloso:2013spa,Kehagias:2013rpa,Horn:2014rta} and show how our results from previous sections apply in this case. 

Just like their inflationary counterpart, the consistency relations for Large Scale Structure can be seen as a consequence of symmetry -- the ability to remove a long mode of the gravitational potential through a change of coordinates. The physical interpretation is particularly clear in the regime when all modes are deep inside the Hubble scale (the long modes only have to be outside the sound horizon, which is practically zero in the late Universe). In this case the homogeneous gradient of the long mode of the gravitational potential is locally equivalent to a homogeneous gravitational field. By the Equivalence Principle we can always choose a free falling reference frame where the effects of such a homogeneous gravitational field are removed. As a consequence, two different solutions -- with and without the long mode -- can be related to each other. This leads to the consistency relations for Large Scale Structure. 

As explained in~\cite{Creminelli:2013mca}, the coordinate transformation that in the non-relativistic limit induces a long wavelength gravitational potential is given by
\be
\eta \rightarrow \eta \;, \qquad x^i \rightarrow x^i  + D(\eta) \partial^i \Phi_L \;,
\ee
where $D(\eta)$ is the growth factor of density fluctuations of the long mode. The consistency relation in the non-relativistic limit which follows from this symmetry is
\be
\label{NRgeneral}
\langle \delta_{\vec q} (\eta) \, \delta_{\vec k_1}(\eta_1) \cdots \delta_{\vec k_n}(\eta_n) \rangle'_{q\to 0} =  - P_\delta(q,\eta) \sum_a \frac{D_\delta(\eta_a)}{D_\delta(\eta)} 
\frac{\vec q \cdot \vec k_a}{q^2} \langle \delta_{\vec k_1}(\eta_1) \cdots \delta_{\vec k_n}(\eta_n) \rangle'  \;.
\ee
This result is obtained assuming that the amplitude of the gravitational field $\partial_i\Phi_L$ is small. However, in the absence of primordial non-Gaussianities, it remains valid even for large gradients as discussed in~\cite{Creminelli:2013mca}. 

We are interested in the case of multiple soft limits. The consistency relation with many soft momenta can be derived straightforwardly. For example, the result for the double-soft limit, using the machinery we have developed, has the following form
\begin{align}
\langle \delta_{\vec q_1} (\eta) & \delta_{\vec q_2} (\eta) \, \delta_{\vec k_1}(\eta_1) \cdots \delta_{\vec k_n}(\eta_n) \rangle'_{q_{1,2}\to 0} = \nonumber \\
& \langle \delta_{\vec q_1} (\eta) \delta_{\vec q_2} (\eta) \delta_{\vec q}(\eta) \rangle' \( - \sum_a \frac{D_\delta(\eta_a)}{D_\delta(\eta)} 
\frac{\vec q \cdot \vec k_a}{q^2} \) \langle \delta_{\vec k_1}(\eta_1) \cdots \delta_{\vec k_n}(\eta_n) \rangle'  \nonumber \\
& + P_\delta(q_1,\eta)P_\delta(q_2,\eta) \(\sum_a \frac{D_\delta(\eta_a)}{D_\delta(\eta)} \frac{\vec q \cdot \vec k_a}{q^2} \sum_b \frac{D_\delta(\eta_b)}{D_\delta(\eta)} \frac{\vec q \cdot \vec k_b}{q^2} \) \langle \delta_{\vec k_1}(\eta_1) \cdots \delta_{\vec k_n}(\eta_n) \rangle' \;.
\end{align}
This expression is analogous to the consistency relation for dilations \eqref{2dilationallN}.
Notice that in this expression we have an additional term compared to~\cite{Creminelli:2013poa}. This is because we didn't assume Gaussian initial conditions. Unfortunately, the contribution of the first term in the expression above seems to be always small. For example, even if we have large equilateral non-Gaussianities (which can be the case in single-field models of inflation with reduced $c_s$), then the relative size of the two terms is proportional to
\be
\frac{\mbox{term 1}}{\mbox{term 2}} \sim \frac{f_{\rm{NL}}^{\rm{eq.}} }{D(\eta)} \frac{q}{k} \;,
\ee
and therefore always parametrically suppressed by $q/k$. Even with optimistic numbers $f_\mathrm{NL}^{\mathrm{eq.}}\sim 100$ and the squeezing $q/k\sim 0.1$, this ratio is at best at the percent level. It is also important to stress that the both terms on the left hand side always vanishes for the equal-time correlation functions. 

One last thing worth stressing is that in the relativistic consistency relation for LSS similar issues discussed in this paper could appear. With one soft external leg this relation reads 
\begin{align}
&\langle \zeta_{\vec q} \, \delta_{\vec k_1}(\eta_1) \cdots \delta_{\vec k_n}(\eta_n) \rangle'_{q\to 0} =  -  P(q)\left[ 3(n-1) + \sum_a \vec k_{a} \cdot \vec \partial_{k_{a}} + \sum_a D_v(\eta_a)(\partial_{\eta_a}-3\mathcal H(\eta_a))  \right. \nonumber \\
& \quad \qquad  \left.  +\sum_a \left( \int^{\eta_a} D_v(\eta)\mathrm d \eta \right) \vec q \cdot \vec k_a +\frac 12 q^iD_i  + \sum_a D_v(\eta_a)(\partial_{\eta_a}-3\mathcal H(\eta_a)) \vec q \cdot \vec \partial_{k_{a}} \right. \nonumber \\
&  \quad \qquad \left. + 6 \sum_a \Omega_{\rm m} (\eta_a) D_v(\eta_a)\mathcal H(\eta_a)\frac{\vec q\cdot \vec k_a}{k_a^2} \left( f_g(\eta_a) + \frac{k_a^2}{3\mathcal H^2(\eta_a)}  \right)^{-1} \right] \langle \delta_{\vec k_1}(\eta_1) \cdots \delta_{\vec k_n}(\eta_n) \rangle' \;,
\end{align}
where $D_v(\eta)$ and $f_g(\eta)$ are the velocity growth rate and growth function. Apart form many terms related to the time evolution, we also have dilation and SCT operators which are identical to those in inflation. Therefore, many of the conclusion from all previous sections about multiple soft limits also apply in this case. For example, presumably the construction of adiabatic modes at higher order in $q$ will again depend on the ordering of coordinate transformations. These questions require further investigation.

\section{Conclusions}
\label{conclu}

In this paper, we have derived novel consistency relations which constrain the form of cosmological correlation functions in the limit where more than one external momentum is taken soft. Like their single-soft cousins, these identities provide model-independent constraints which must be satisfied by any single-field model of inflation (with the same assumptions as in the single soft case). In this way, they provide robust null tests of the inflationary paradigm, and any violation of them would be extremely interesting. Additionally, we have commented on the application of the same ideas to models of large scale structure and considered consistency relations in this case. Looking forward, there are various further avenues to explore:

\begin{itemize}
\item Here we have worked to lowest order in tensor perturbations. In this sense, our results only hold in models where tensors are negligible (for example in the decoupling limit of small $c_s$ models). However, it is straightforward to generalize our formulae to include tensor perturbations. In particular it would be interesting to work out the consequences of having two soft gravitons in an inflationary correlation function. These questions are most easily approached in the 1PI vertex identity language (though it would also be instructive to understand things from the background wave perspective); it is straightforward to incorporate tensor transformations into the formalism of~\cite{Goldberger:2013rsa} and derive the 1PI Ward identities involving tensors. It should then be possible to derive the higher-soft analogue of the consistency relations of~\cite{Hinterbichler:2013dpa,Berezhiani:2013ewa} order-by-order in $q$. This would be the logical marriage of~\cite{Hinterbichler:2013dpa} and the present work.

\item We presented explicitly the generalization to $N$ soft legs at lowest order in the soft momentum (corresponding to $N$ dilations). It would be interesting to work out systematically the higher-order in $q$ corrections to this identity. At ${\cal O}(q)$, we do not anticipate any difficultly, aside from things being technically complicated. However at ${\cal O}(q^2)$, there should again be subtleties associated to curvature effects, which may be quite interesting.

\item We have focused on multiple-soft limits in inflation, but it would be interesting to derive the consequences of higher soft limits in alternative theories as well. In particular the conformal mechanism for generating inflationary perturbations~\cite{Rubakov:2009np,Libanov:2010nk,Creminelli:2010ba,Hinterbichler:2011qk,Hinterbichler:2012mv,Creminelli:2012my,Hinterbichler:2012yn,Hinterbichler:2012fr}, for which the single-soft consistency relations have recently been derived~\cite{Creminelli:2012qr}.

\item In our discussion of multiple soft limits in large scale structure, we have worked in the Newtonian approximation. It would be interesting to go beyond this approximation and investigate the double-soft relations in the fully relativistic regime. Similar to the primordial case, these relations involve the action of dilation and conformal symmetries, with nontrivial commutators~\cite{Creminelli:2013mca}. We therefore expect that many of the same subtleties should arise. 

\end{itemize}

\vspace{.4cm}
 \noindent
{\bf Note:} While finishing this paper, we became aware of~\cite{Mirbabayi:2014zpa}, which also considers double soft relations in inflation. Where our results overlap, we agree. We thank the authors for sharing an early draft with us.

\vspace{.4cm}
\noindent
{\bf Acknowledgements:} We thank Lasha Berezhiani, Paolo Creminelli, Mehrdad Mirbabayi and Junpu Wang for helpful discussions. We would particularly like to thank Yi Wang for sharing with us his Mathematica notebook for calculating the 4-point function in models with small speed of sound. This work is supported in part by NASA ATP grant NNX11AI95G and NSF CAREER Award PHY-1145525 (J.K.); the Kavli Institute for Cosmological Physics at the University of Chicago through grant NSF PHY-1125897, an endowment from the Kavli Foundation and its founder Fred Kavli and by the Robert R. McCormick Postdoctoral Fellowship (A.J.).

\appendix
\addcontentsline{toc}{section}{Appendix}

\section*{Appendix: 4-Point Function in models with small sound speed}
\label{4pointappendix}
\renewcommand{\theequation}{A-\arabic{equation}}
\setcounter{equation}{0} 
\setcounter{subsection}{0} 
\renewcommand{\thesubsection}{A-\Roman{subsection}}
Here we collect expressions from~\cite{Huang:2006eha,Chen:2009bc} necessary to perform the check of the double-soft Ward identity conducted in Sec.~\ref{corrcheck}. We are interested in models with small speed of sound, these descend from an action of the form:
\be
S = \int\rd^4x\sqrt{-g}\left(\frac{M_{\rm Pl}^2}{2}R +P(X, \phi)\right)~,
\ee
where $X = -\frac{1}{2}(\partial\phi)^2$. Expanding about an arbitrary time-dependent background $\phi = \bar\phi(t)+\vp$ and an FLRW background we obtain~\cite{Huang:2006eha,Chen:2006nt}
\begin{align}
\nonumber
S = \int\rd^4x~a^3\bigg[ &\frac{1}{2}\left(P_{,X}-P_{,XX}\dot{\bar\phi}^2\right)\dot\vp^2 -\frac{1}{2a^2}P_{,X}(\vec\vp)^2\\
&+\left(\frac{1}{2}P_{,XX}\dot{\bar\phi}+\frac{1}{6}P_{,XXX}\dot{\bar\phi}^3\right)\dot\vp^3-\frac{1}{2a^2}P_{,XX}\dot{\bar\phi}~\dot\vp(\vec\nabla\vp)^2\\\nonumber
&+\frac{1}{24}P_{,XXXX}\dot{\bar\phi}^4~\dot\vp^4+\frac{1}{4}P_{,XXX}\dot{\bar\phi}^2~\dot\vp^2\left(\dot\vp^2-\frac{1}{a^2}(\vec\nabla\vp)^2\right)+\frac{1}{8}P_{,XX}\left(\dot\vp^2-\frac{1}{a^2}(\vec\nabla\vp)^2\right)^2\bigg]~.
\end{align}
If we make the following definitions
\begin{align}
\nonumber
&XP_{,X} = \epsilon H^2~,~~~~~~~~~~~~~~~~~~~~~~~~~~~~~~~~~~\lambda= X^2P_{,XX}+\frac{2}{3}X^3P_{,XXX} ~,\\
& \Sigma =XP_{,X}+2X^2P_{,XX}= \frac{\epsilon H^2}{c_s^2}~,~~~~~~~~~~~ \mu=\frac{1}{2}X^2P_{,XX}+2X^3P_{,XXX}+\frac{2}{3}X^4P_{,XXXX}~,
\end{align}
and identify
\be
\zeta = \frac{H}{\dot{\bar\phi}}\vp~,
\ee
the action takes the form~\cite{Chen:2009bc}
\begin{align}
\nonumber
S = \int\rd^4x~a^3\bigg[ &\frac{\epsilon}{c_s^2}\left(\dot\zeta^2 - \frac{1}{a^2}(\vec\nabla\zeta)^2\right)\\\nonumber
&-2\frac{\lambda}{H^3}\dot\zeta^3+\frac{1}{a^2}\frac{\Sigma}{H^3}(1-c_s^2)\dot\zeta(\vec\nabla\zeta)^2\\
&+\frac{\mu}{H^4}\dot\zeta^4-\frac{1}{a^2H^4}(3\lambda-\Sigma(1-c_s^2))\dot\zeta^2(\vec\nabla\zeta)^2+\frac{1}{4a^4H^4}\Sigma(1-c_s^2)(\vec\nabla\zeta)^4\bigg]
\end{align}
The two-point function is given by
\be
\langle\zeta_{\vec k}\zeta_{\vec k'}\rangle = (2\pi)^3\delta^{(3)}(\vec k+\vec k') \frac{H^2}{4M_{\rm Pl}^2 c_s\epsilon}\frac{1}{k^3} \equiv (2\pi)^3\delta^{(3)}(\vec k+\vec k')P(k)~.
\ee
Now, we specialize to the case where $\lambda = \mu = 0$. This is a consistent choice---from the viewpoint of checking the consistency relations, we need not be fully general. The interaction Hamitonian is given by (we now work in conformal time $a\rd\eta = \rd t$)\footnote{Note that transforming from the interaction Lagrangian to the interaction Hamiltonian is non-trivial at this order.}
\begin{align}
{\cal H}^{(3)}_{\rm int.} &= \rd^3x\rd\eta \left(-\frac{a}{H^3}\Sigma(1-c_s^2)\zeta'(\vec\nabla\zeta)^2\right)\\
{\cal H}^{(4)}_{\rm int.} &= \rd^3x\rd\eta \frac{\Sigma(1-c_s^2)}{H^4}\left(-\zeta'^2(\vec\nabla\zeta)^2-\frac{1}{2}c_s^2(\vec\nabla)^4\right)~.
\end{align}
The three point function is given by
\be
\langle\zeta_{\vec k_1}\zeta_{\vec k_2}\zeta_{\vec k_3}\rangle = -i\int_{-\infty}^{t}\rd t'\langle[\zeta_{\vec k_1}\zeta_{\vec k_2}\zeta_{\vec k_3}(t), {\cal H}^{(3)}_{\rm int.}(t')]\rangle~.
\ee
Evaluating the in-in integral, we obtain~\cite{Chen:2009bc,Chen:2006nt}
\begin{align}
\langle\zeta_{\vec k_1}\zeta_{\vec k_2}\zeta_{\vec k_3}\rangle' =~& P(k_2)P(k_3)\frac{1}{2 k_1^3} \bigg[ \left(\frac{1}{c_s^2}-1\right)\frac{12 k_1^2k_2^2k_3^2}{k_t^3}  \nonumber \\
& + \left( \frac{1}{c_s^2}-1 \right) \bigg( -\frac{8}{k_t}\sum_{i>j} k_i^2k_j^2 +\frac{4}{k_t^2}\sum_{i\neq j}k_i^2k_j^3 +\sum_i k_i^3  \bigg) \bigg]~.
\label{3ptsmallcs}
\end{align}
Similarly, the equal-time 4-point function is given by in the in-in formalism
\begin{align}
\nonumber
\langle\zeta_{\vec k_1}\zeta_{\vec k_2}\zeta_{\vec k_3}\zeta_{\vec k_4}\rangle &= -i\int_{-\infty}^{t}\rd t'\langle[\zeta_{\vec k_1}\zeta_{\vec k_2}\zeta_{\vec k_3}\zeta_{\vec k_4}(t), {\cal H}^{(4)}_{\rm int.}(t')]\rangle\\
&~~+ \int_{-\infty}^t\rd t'\int_{-\infty}^t\rd t''\langle  {\cal H}_{\rm int.}^{(3)}(t')\zeta_{\vec k_1}\zeta_{\vec k_2}\zeta_{\vec k_3}\zeta_{\vec k_4}(t){\cal H}_{\rm int.}^{(3)}(t'')\rangle\\
\nonumber
&~~-2{\rm Re}\left(\int_{-\infty}^t\rd t'\int_{-\infty}^{t'}\rd t''\langle\zeta_{\vec k_1}\zeta_{\vec k_2}\zeta_{\vec k_3}\zeta_{\vec k_4}(t){\cal H}_{\rm int.}^{(3)}(t'){\cal H}_{\rm int.}^{(3)}(t'')\rangle\right)~.
\end{align}
Computing this is an extremely intricate task, worked out carefully in~\cite{Chen:2009bc}, whose notation we adopt for convenience. The final answer is:
\be
\langle\zeta_{\vec k_1}\zeta_{\vec k_2}\zeta_{\vec k_3}\zeta_{\vec k_4}\rangle' =P(k_1)P(k_2)P(k_3)\frac{2^3}{k_4^3}{\cal T}(k_1, k_2, k_3, k_4)~,
\label{4pointzetas}
\ee
where the shape function is given by~\cite{Chen:2009bc}
\be
{\cal T}(k_1, k_2, k_3, k_4) = \left(\frac{1}{c_s^2}-1\right)^2 T_{s3}-\left(\frac{1}{c_s^2}-1\right)T_{c2}+\left(\frac{1}{c_s^2}-1\right)T_{c3}~.
\ee
The two contributions from the contact interaction are given by~\cite{Chen:2009bc}
\begin{align}
T_{c2} &= \frac{1}{8}\left(\frac{1}{c_s^2}-1\right)\frac{k_1^2k_2^2 (\vec k_3\cdot\vec k_4)}{K^3}\left(1+\frac{3(k_3+k_4)}{K}+\frac{12_3k_4}{^2}\right)+23~{\rm perms.}\\
T_{c3} &= \frac{1}{32}\left(\frac{1}{c_s^2}-1\right)\frac{(\vec k_1\cdot\vec k_2)(\vec k_3\cdot\vec k_4)}{K}\left(1+\frac{\sum_{i<j}k_ik_j}{K^2}+\frac{3k_1k_2k_3k_4}{K^3}\sum_{i=1}^4\frac{1}{k_i}+\frac{12k_1k_2k_3k_4}{K^4}\right)+23~{\rm perms.}
\end{align}
The contribution from the exchange diagram is more complicated, the final result is~\cite{Chen:2009bc}\footnote{When performing the sum over permutations of the external momenta, $M$ is left invariant.}
\begin{align}
\nonumber
 T_{s3} = \left(\frac{1}{c_s^2}-1\right)^2\bigg[&\frac{1}{2^7} (\vec k_1\cdot\vec k_2)(\vec k_3\cdot\vec k_4)k_{12}F(k_1, k_2, k_1+k_2+k_{12})F(k_3, k_4, M)\\\nonumber
 &+\frac{1}{2^5}(\vec k_1\cdot \vec k_2)(\vec k_{12}\cdot \vec k_4)\frac{k_3^2}{k_{12}}F(k_1, k_2, k_1+k_2+k_{12})F(k_{12}, k_4, M)\\\nonumber
 &-\frac{1}{2^5}(\vec k_{12}\cdot\vec k_2)(\vec k_{12}\cdot \vec k_4)\frac{k_1^2k_3^2}{k_{12}^3} F(k_{12}, k_2, k_1+k_2+k_{12})F(k_{12}, k_4, M)\\\nonumber
& +\frac{1}{2^6}(\vec k_1\cdot\vec k_2)(\vec k_3\cdot\vec k_4) k_{12} G_{bb}(k_1, k_2, k_3, k_4)\\\nonumber
 &+\frac{1}{2^5}(\vec k_1\cdot\vec k_2)(\vec k_{12}\cdot \vec k_4)\frac{k_3^2}{k_{12}} G_{bb}(k_1, k_2, k_{12}, k_4)\\\nonumber
& -\frac{1}{2^5}(\vec k_{12}\cdot\vec k_2)(\vec k_{3}\cdot \vec k_4)\frac{k_1^2}{k_{12}} G_{bb}(-k_{12}, k_2, k_3, k_4)\\
& -\frac{1}{2^4}(\vec k_{12}\cdot\vec k_2)(\vec k_{12}\cdot \vec k_4)\frac{k_1^2k_3^2}{k_{12}^3} G_{bb}(-k_{12}, k_2, k_{12}, k_4)\bigg] +23~{\rm perms.}~,
\end{align}
where $\vec k_{12} = \vec k_1+\vec k_2$; $M = k_3+k_4+k_{12}$ and $K = k_1+k_2+k_3+k_4$ and the functions are given by~\cite{Chen:2009bc}
\begin{align}
F(\alpha_1, \alpha_2, ) = &\frac{1}{m^3}\left(2\alpha_1\alpha_2+(\alpha_1+\alpha_2)m+m^2\right)\\\nonumber
G_{bb}(\alpha_1, \alpha_2, \alpha_3, \alpha_4) = &\frac{1}{M^3K}\left(2\alpha_3\alpha_4+(\alpha_3+\alpha_4)M+M^2\right)\\\nonumber
&+\frac{1}{M^3K^2}\left(2\alpha_3\alpha_4(\alpha_1+\alpha_2)+\Big(2\alpha_3\alpha_4+(\alpha_1+\alpha_2)(\alpha_3+\alpha_4)\Big)M+\sum_{i=1}^4\alpha_i M^2\right)\\\nonumber
&+\frac{2}{M^3K^3}\left(2\prod_{i=1}^4\alpha_i+\Big(2\alpha_3\alpha_4(\alpha_1+\alpha_2)+\alpha_1\alpha_2(\alpha_3+\alpha_4)\Big)M+\sum_{i<j}\alpha_i\alpha_jM^2\right)\\
&+\frac{6}{M^2K^4} \prod_{i=1}^4\alpha_i\left(2+M\sum_{i=1}^4\frac{1}{\alpha_i}\right)+\frac{24}{MK^5}\prod_{i=1}^4\alpha_i~.
\end{align}
Putting all of this together, we can compute the double squeezed limit of~\eqref{4pointzetas}:
\begin{align}
\nonumber
\lim_{q_1,q_2\to 0}\bigg(
\frac{\langle\zeta_{\vec q_1}\zeta_{\vec q_2}\zeta_{\vec k_1}\zeta_{\vec k_2}\rangle'}{P(q_1)P(q_2)}\bigg) =~ &P(k)\left(1-\frac{1}{c_s^2}\right)\left(\frac{\langle\zeta_{\vec q_1}\zeta_{\vec q_2}\zeta_{-\vec q}\rangle}{P(q_1)P(q_2)}\right)\left(2\frac{q^2}{k^2}-\frac{5}{4}\frac{(\vec k\cdot\vec q)^2}{k^4}\right)\\
&+P(k)\left(1-\frac{1}{c_s^2}\right)\left(\frac{3}{4}\frac{(\vec q_1\cdot\vec q_2)}{k^2}+\frac{5}{2}\frac{(\vec k\cdot \vec q_1)(\vec k\cdot\vec q_2)}{k^4}\right)~,
\end{align}
where $\langle\zeta_{\vec q_1}\zeta_{\vec q_2}\zeta_{-q}\rangle$ is given by~\eqref{3ptsmallcs}.

\renewcommand{\em}{}
\bibliographystyle{utphys}
\addcontentsline{toc}{section}{References}
\bibliography{highersoft20}

\end{document}